\documentclass{article}

\usepackage[T1]{fontenc}

\usepackage{microtype}
\usepackage{graphicx}
\usepackage[export]{adjustbox}
\usepackage{subfig}
\usepackage[strings]{underscore}
\usepackage{amsfonts}
\usepackage{algorithm}
\usepackage{algorithmic}
\usepackage{bbm}
\usepackage{dsfont}
\usepackage{amsmath}

\usepackage{hyperref}

\def\EQ#1{\begin{eqnarray}#1\end{eqnarray}}

\usepackage[accepted]{icml2021}

\usepackage{microtype}
\usepackage{mathtools}
\usepackage{amssymb}
\usepackage{physics}

\begin{document}

\twocolumn[
\icmltitle{Unsupervised strategies for identifying optimal parameters in Quantum Approximate Optimization Algorithm}

\begin{icmlauthorlist}
\icmlauthor{Charles Moussa}{liacs}
\icmlauthor{Hao Wang}{liacs}
\icmlauthor{Thomas B\"{a}ck}{liacs}
\icmlauthor{Vedran Dunjko}{liacs}

\end{icmlauthorlist}

\icmlaffiliation{liacs}{LIACS, Leiden University, Niels Bohrweg 1, 2333 CA Leiden, Netherlands}

\icmlcorrespondingauthor{Charles Moussa}{c.moussa@liacs.leidenuniv.nl}
\icmlcorrespondingauthor{Vedran Dunjko}{v.dunjko@liacs.leidenuniv.nl}

\icmlkeywords{Clustering, Quantum Computing, QAOA}

\vskip 0.3in
]

\printAffiliationsAndNotice{}  

\begin{abstract} % abstract
As combinatorial optimization is one of the main quantum computing applications, many methods based on parameterized quantum circuits are being developed. In general, a set of parameters are being tweaked to optimize a cost function out of the quantum circuit output. One of these algorithms, the Quantum Approximate Optimization Algorithm stands out as a promising approach to tackling combinatorial problems. However, finding the appropriate parameters is a difficult task. Although QAOA exhibits concentration properties, they can depend on instances characteristics that may not be easy to identify, but may nonetheless offer useful information to find good parameters. In this work, we study unsupervised Machine Learning approaches for setting these parameters without optimization. We perform clustering with the angle values but also instances encodings (using instance features or the output of a variational graph autoencoder), and compare different approaches. These angle-finding strategies can be used to reduce calls to quantum circuits when leveraging QAOA as a subroutine. We showcase them within Recursive-QAOA up to depth $3$ where the number of QAOA parameters used per iteration is limited to $3$, achieving a median approximation ratio of $0.94$ for MaxCut over $200$ Erd\H{o}s-R\'{e}nyi graphs. We obtain similar performances to the case where we extensively optimize the angles, hence saving numerous circuit calls.
\end{abstract}

\section{Introduction}
\label{intro}

\par
Noisy Intermediate-Scale Quantum (NISQ) era hardware \cite{Preskill2018quantumcomputingin} faces many limiting challenges preventing fault-tolerant quantum algorithm execution (e.g., the number of qubits, decoherence, etc.). Hence near-term hybrid quantum-classical algorithms were designed as an alternative for applications such as quantum chemistry problems \cite{VQE}, quantum machine learning \cite{PQC} and combinatorial optimization \cite{QAOA}. 
\par 
With a user-specified depth $p$, the Quantum Approximate Optimization Algorithm (QAOA) \cite{QAOA} consists of a quantum circuit involving $2p$ real parameters (or angles). QAOA exhibits a few properties that makes it interesting for combinatorial optimization such as a perfect theoretical performance at infinite depth \cite{QAOA}, a sampling advantage \cite{qaoaSupremacy} and the concentration of parameters \cite{qaoaConcentrates}. The latter suggests that optimal parameters found for one instance can be reused on another. Most importantly, this means we can reduce the classical optimization loop and number of calls to a quantum device (saving runtime of QAOA-featured algorithms). 
\par 
Many works have studied or illustrated this concentration property \cite{qaoaConcentrates, qaoaperf,Khairy2020,9605323,qaoatransfer,flipe,paramconcentrationfolklore,streiftrain, crooks}. However, in many algorithms which feature QAOA as a subroutine \cite{daqqaoa,daqqaoa2, Moussa2021,qls,rqaoa}, many distributions of instances are generated and several areas of parameter concentrations may arise. Hence, balancing between finding good QAOA parameters and reducing circuit calls will be key to QAOA-featured algorithms.
\par
In this work, we propose to apply unsupervised learning for setting QAOA angles, namely clustering. Our main contributions are as follows:
\begin{itemize}
\item We consider different approaches for the problem of setting QAOA angles with clustering: using directly the angle values, instance features, and the output of a variational graph autoencoder as input to the clustering algorithm.
\item We analyze our methods by comparing them on two types of problems: MaxCut on Erd\H{o}s-R\'{e}nyi graphs and Quadratic Unconstrained Binary Problems on random dense matrices. 
\item We demonstrate that our techniques can be used to learn to set QAOA parameters with respectively a less than $1-2\%$ reduction (in relative value) in approximation ratio in cross-validation while reducing circuit calls. 
\item We show that leveraging instance encodings for angle setting strategies yields better results than using angle values only.
\item Finally, we demonstrate their usage in Recursive-QAOA (RQAOA) \cite{rqaoa} up to depth $3$ on the Erd\H{o}s-R\'{e}nyi graphs. We limit the number of QAOA circuit calls per iteration to $3$ (in contrast to a de novo optimization which would require many more calls), and achieve a $0.94$ median approximation ratio. With our approaches, we obtain similar performances to the case where we extensively optimize the angles, hence saving numerous circuit calls.
\end{itemize}
\par 
\vspace{1mm} The structure of the paper is as follows. Section~\ref{generalities} provides the necessary background and related works. Section~\ref{revisitconcentration} analyses the optimal angles found in both problems, pointing to concentration effects and the suitability of clustering. Section~\ref{clustering} shows different unsupervised learning strategies using different data encoding for clustering and the comparison between them. Section~\ref{resrqaoa} sums up our experiments on RQAOA. We conclude this work with a discussion in Section~\ref{conclusion}. 

\section{Background}
\label{generalities}

\subsection{QUBO and QAOA} Quadratic Unconstrained Binary Optimization (QUBO) problems are specified by the formulation $ \min_{x \in \{0,1\}^n} \\\ \sum_{i \le j} x_i Q_{ij} x_j$ where $n$ is the dimensionality of the problem and $Q\in\mathbb{R}^{n\times n}$.
This formulation is connected to the task of finding so called \guillemotleft ground states\guillemotright of \guillemotleft Ising models\guillemotright, i.e., configurations of binary labels $\{1,-1\}$ minimising the energy of spin Hamiltonians, commonly tackled in statistical physics and quantum computing, i.e.,:
\EQ{
 \textstyle\min_{s \in \{-1,1\}^n} \sum_i h_i s_i + \textstyle\sum_{j>i} J_{ij} s_i s_j, \label{ising}
 }
where $h_i$ are the biases and $J_{ij}$ the interactions between spins.
QUBO can express an exceptional variety of combinatorial optimization (CO) problems such as Quadratic Assignment, Constraint Satisfaction Problems, Graph Coloring, and Maximum Cut \cite{Kochenberger2006}.
\par 
The QAOA algorithm \cite{QAOA} was inspired by adiabatic quantum computing with the goal to tackle CO problems. It consists of a quantum circuit whose construction depends on the classical cost function. Indeed, the latter is encoded in a quantum Hamiltonian defined on $N$ qubits by replacing each variable $s_i$ in Eq.~\eqref{ising} by the single-qubit operator $\sigma_i^z$:
\EQ{ H_C =  \textstyle\sum_i h_i \sigma_i^z + \sum_{j>i} J_{ij}  \sigma_i^z \sigma_j^z .}
Here, the bitstring corresponding to the ground state of $H_C$ also minimizes the cost function. Another Hamiltonian named \emph{mixer} $  H_B = \sum_{j=1}^N  \sigma_j^x$ is also employed in QAOA. These operators are then used for building a quantum circuit with real parameters and organized as layers. This circuit is initialized in the $\ket{+}^{\otimes N}$ state, corresponding to all bitstrings in superposition with equal probability of being measured. Then, applying $p$ layers sequentially yields the following quantum state:
$$  \ket{\mathbf{\gamma},\mathbf{\beta}} = e^{-i\beta_p H_B} e^{-i\gamma_p H_C} \cdots e^{-i\beta_1 H_B} e^{-i\gamma_1 H_C} \ket{+}^{\otimes N},  $$
defined by $2p$ real parameters $\gamma_i,\beta_i, i=1...p$ or \emph{QAOA angles} as they correspond to angles of parameterized quantum gates. Such output corresponds to a probability distribution over all possible bitstrings. The classical optimization challenge of QAOA is to find the sequence of angles $\mathbf{\gamma}, \mathbf{\beta}$ minimizing the expected value of the cost function from the measurement outcome. In the limit of infinite depth, the distribution will converge to the global optimum. 
\par 
An interesting property of the algorithm is the concentration of the QAOA objective for fixed angles \cite{qaoaConcentrates} due to typical instances having (nearly) the same value of the objective function. Additionally, the QAOA landscape is instance-independent when instances come from a \guillemotleft reasonable\guillemotright~distribution (with the number of certain types of subgraphs of fixed size themselves concentrate, which in turn implies the values concentrate). Hence, we can focus on finding good parameters on a subset of instances that could be re-applied to new ones, with a few extra calls to the quantum device in order to refine. As stated earlier, in the most general case, characterizing distributions which are \guillemotleft reasonable\guillemotright~may be involved, or even characterizing the distribution at hand may be hard. Previous work \cite{qaoaConcentrates, streiftrain, crooks} referenced \cite{paramconcentrationfolklore} reported concentrations over optimal parameters even when QAOA is applied on random instances. These distributions over optimal parameters are empirically shown to behave non-trivially with respect to $n$.  \cite{paramconcentrationfolklore} pointed out this problem as \guillemotleft folklore of concentrations\guillemotright~.
\par 
Hence, even though angles concentrate in many settings asymptotically, for finite-size problems, different areas of concentration may rise. Therefore, choosing good angle values is challenging, especially when considering the runtime of quantum algorithms. As such, some studies built on this property and resorted to using Machine Learning (ML) or characterizing instances by some properties for finding good QAOA parameters. We present a few of them in the next subsection.

\subsection{Related Work}

Many previous works have extensively employed the concentration property \cite{qaoaperf, Khairy2020, 9605323, qaoatransfer, flipe, paramconcentrationfolklore, streiftrain, crooks}. Among them, a few employed ML or designed strategies for setting good QAOA parameters for different objectives. 
In \cite{Khairy2020}, a simple kernel density model was trained on the best angles and instances solved by QAOA to exhibit better QAOA optimization than the Nelder-Mead optimizer. Parameter fixing strategies for QAOA are also studied in \cite{9605323, qaoaperf} where the best-found angles at depth $p$ are used as starting points for depth $p+1$ before using a classical optimizer. 
\par 
\cite{streiftrain} present a strategy to find good parameters for QAOA based on topological properties of the problem graph and tensor network techniques. \cite{qaoatransfer} point out that the success of transferability of parameters between different problem instances can be explained and predicted based on the types of subgraphs composing a graph. 
Finally, meta-learning is used in \cite{flipe} to learn good initial angles for QAOA. They focused on initialization-based meta-learners in which a single set of parameters is used for a distribution of problems as initial parameters of a gradient-based optimizer. The meta-learner is a simple neural network that takes as inputs some meta-features of the QAOA circuit to predict the angles to apply (depth and which angle to output the value). However, no instance-related features are involved in their work.
\par
In our case, we focus on clustering with the goal of proposing many parameter values to try for new QAOA circuits. In contrast to all the approaches we discussed above, we do not use a classical optimization loop after setting them. Hence, our approaches allow balancing between circuit calls of small quantum computers and performances. Such settings for instance naturally occur in divide-and-conquer-type schemes to enable smaller quantum computers to improve optimization \cite{daqqaoa,daqqaoa2, Moussa2021,qls}, or in Recursive-QAOA \cite{rqaoa} as we demonstrate later.

\section{Revisiting the concentration property}
\label{revisitconcentration}

In contrast to previous related works, we propose unsupervised approaches that also exploit these concentration effects. We take a data-driven approach where from examples of good angles, we will infer new good angles for new instances. Namely, we use clustering in order to obtain clusters that can be used to reduce calls to the quantum device to small numbers (in our case, less than 10) when applying QAOA on new instances, without further optimization.
\par 
We take a usual ML approach to this problem. First, from generated instances, we apply \emph{exploratory data analysis} \cite{edaref} (EDA) that suggests clustering may be a good approach for recommending good angles to new instances. Namely, we look at the density of angle values and apply t-distributed stochastic neighbor embedding (t-SNE) \cite{tsne} for visualizing concentration effects. t-SNE is a nonlinear dimensionality reduction technique for mapping high-dimensional data to a lower $d$-dimensional space (typically $d\in\{2,3\}$). Briefly, this method constructs a probability distribution to measure the similarity between each pair of points, where closer pairs are assigned with a higher probability. Then, in the lower-dimensional space $\mathbb{R}^d$, we use a Student $t$-based distribution to quantify the similarity among the embeddings of the original data points. Finally, the optimal embeddings are chosen by minimizing the Kullback–Leibler divergence between the similarity distributions in the original and the lower-dimensional spaces. We follow by explaining how clustering is used in order to recommend angles for new instances. The approaches we outline differ in input to the clustering algorithm. We consider clustering from the angle values directly but also from instance encodings. Finally, we compare these approaches allowing us to provide recommendations for their usage.

\subsection{Data generation}

We generated two datasets that show different concentration behavior. The first one consists of $200$ Erd\H{o}s-R\'{e}nyi graphs for MaxCut problems. The graphs have $10, 12, 14, 16$ and $18$ nodes. We utilized the following probabilities of edge creation: $0.5, 0.6, 0.7, 0.8$. We have generated $10$ graphs per number of nodes and probability. The second dataset consists of $100$ instances of QUBO problems, specified by their weight matrix $Q$ ($20$ per aforementioned number of nodes). Their coefficients are sampled uniformly in $[-1,1]$. For the purpose of computing approximation ratios, we are interested in $C_{opt}$ -- the maximal value of the MaxCut (or QUBO) -- over all possible bit configurations, and as a reference, this was computed using brute-force. Our experiments were achieved using a classical simulator.
\par 
We then obtained for each problem the best set of angles by running the BFGS optimizer \cite{bfgs} $1000$ times for $p=1,2,3$, and selecting the ones which achieve the best QAOA objective. BFGS with random restarts is deemed a very good optimizer for continuous differentiable functions~\cite{HansenARFP10}. These angles are saved as a database and apply unsupervised approaches to learn to set optimal angles for unseen instances. Our approach is clearly optimization method-specific but can be applied to other state-of-the-art optimizers. Different optimizers would give different data (as the optimizers could fail to find the optimal QAOA parameters) but they can be combined and one would select the best set of angles found among all considered.

\subsection{Exploratory Data Analysis}
\label{eda}
 
\begin{figure}[!ht]
\centering
\includegraphics[width=0.45\textwidth]{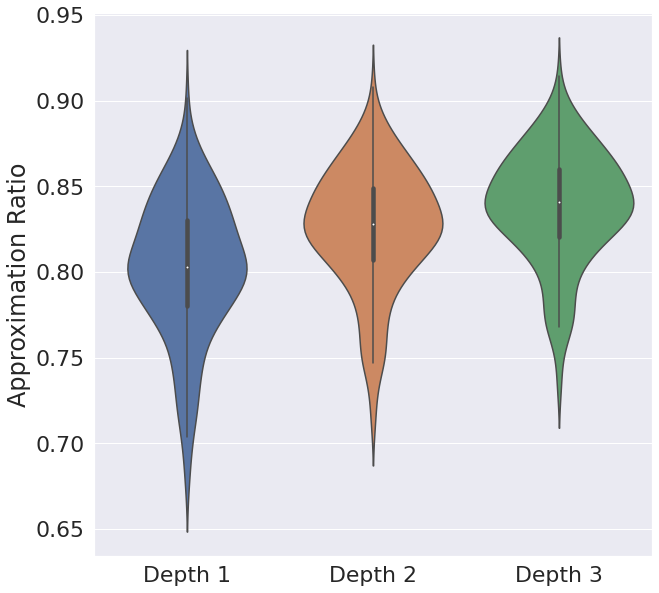}
\includegraphics[width=0.45\textwidth]{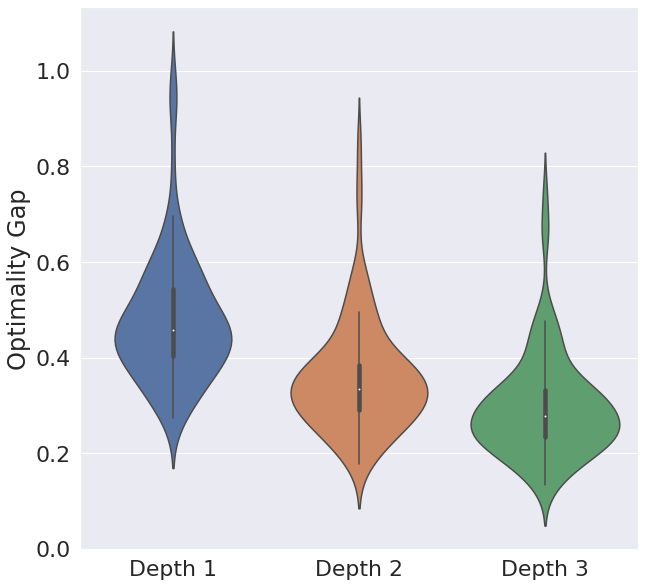}
\caption{
\small Violin plots of ratios on MaxCut and optimality gaps over QUBOs (bottom plot) for $p = 1, 2, 3$. The respective median by depth is $0.802954, 0.827901, 0.840478$ for MaxCuts and $0.457434, 0.335144, 0.278984 $ for QUBOs, illustrating improved performances with increased depth.}
\label{ratios}
\end{figure}

Having obtained the optimal angles, we apply EDA to observe concentration effects. We look at their corresponding performance ratios using the average cost yielded by QAOA for angles $\gamma, \beta$ denoted with $E_{\gamma,\beta} (C)$. For MaxCut on unweighted Erd\H{o}s-R\'{e}nyi graphs, we compute approximation ratios as $ \frac{E_{\gamma,\beta} (C)}{C_{opt}}$. This value is upper bounded by 1, which is the optimal value. For QUBOs, we compute optimality gaps $ \frac{C_{opt} - E_{\gamma,\beta} (C)}{C_{opt}} $ as the optima were all negative and the closer to 0, the better. We show boxplots in Fig~\ref{ratios} the ratios wrt depth. Increasing depth results in better ratios. 

\begin{figure*}[!ht]
\centering
\subfloat[]{\includegraphics[width=0.25\textwidth]{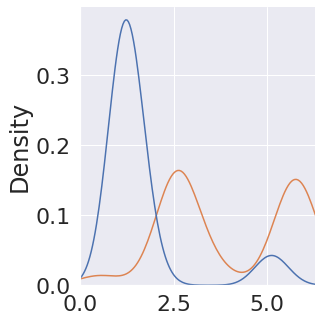}}
\subfloat[]{\includegraphics[width=0.25\textwidth]{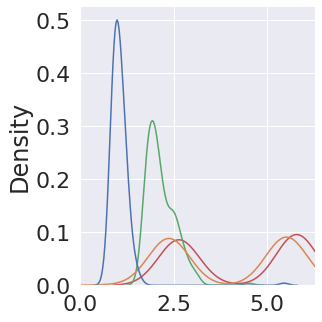}}
\subfloat[]{\includegraphics[width=0.35\textwidth]{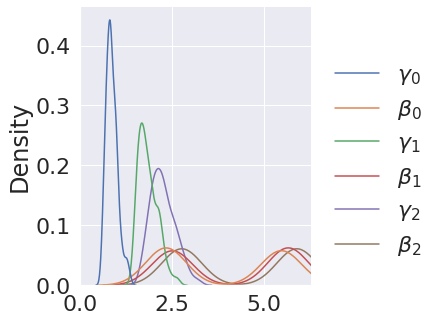}}
\newline
\subfloat[]{\includegraphics[width=0.25\textwidth]{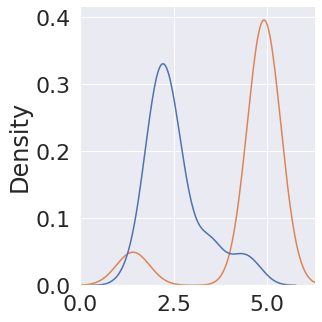}}
\subfloat[]{\includegraphics[width=0.25\textwidth]{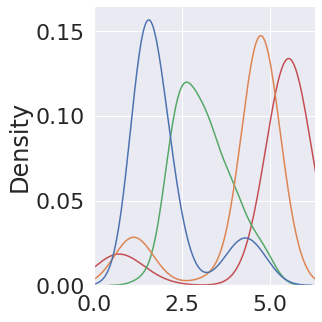}}
\subfloat[]{\includegraphics[width=0.35\textwidth]{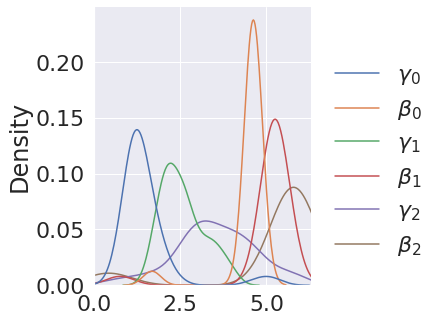}}
\caption{
\small Distribution of angle values $\gamma_i, \beta_i$ for each depth. Plots a), b) and c) concern MaxCut problems while the others refer to the dense QUBO matrices. We witness concentration effects of the angle values, suggesting the suitability of clustering as an angle setting strategy.}
\label{densityangles}
\end{figure*}

Next, we looked at the distribution of $\gamma_i, \beta_i$ values. Fig~\ref{densityangles} shows that the concentration per each parameter is significant since their corresponding density functions are quite peaky. Also, we also observed multiple clusters of angles as the density functions are multimodal. Finally, we applied t-SNE with two components to visualize the angle values in $2D$ for $p=2,3$. This highlights potentially a number of clusters for each depth and problem. Note that it may be possible that we may not obtain global optima with these angles, or know if they are unique.
\par 
We notice that the probability of edge creation, represented by a different color, does not seem to influence the clusters. For dense QUBOs, we observe one important cluster and a few instances that start to form another. Finally, in the dense instances case, we witness a more important spread in angle values at depth $1$. This can be explained by differences between instances. Although the concentration effect is present, such order of magnitude will impact the performances of parameter setting strategies, and make an interesting playground to benchmark them. 
\par
Using clustering techniques can then reveal potential areas of QAOA angles values where good angles can be found to try on new instances. The angle values related to clusters can be used as recommendations for new instances. This becomes interesting as this enables lowering runtime and allow comparing based on function evaluations, or on the number of quantum circuit calls, in algorithms where QAOA would be used as a subroutine.

\begin{figure*}[!ht]
\centering
\subfloat[]{\includegraphics[width=0.3\textwidth]{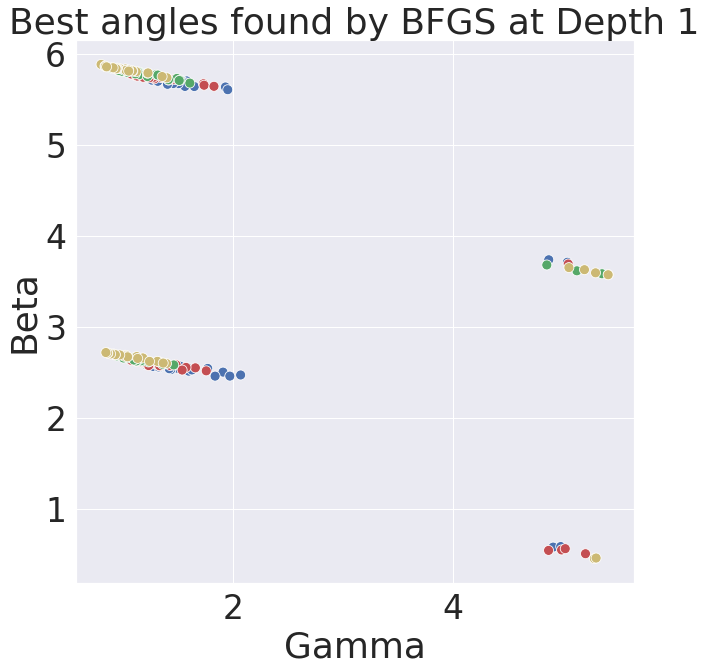}}
\subfloat[]{\includegraphics[width=0.3\textwidth]{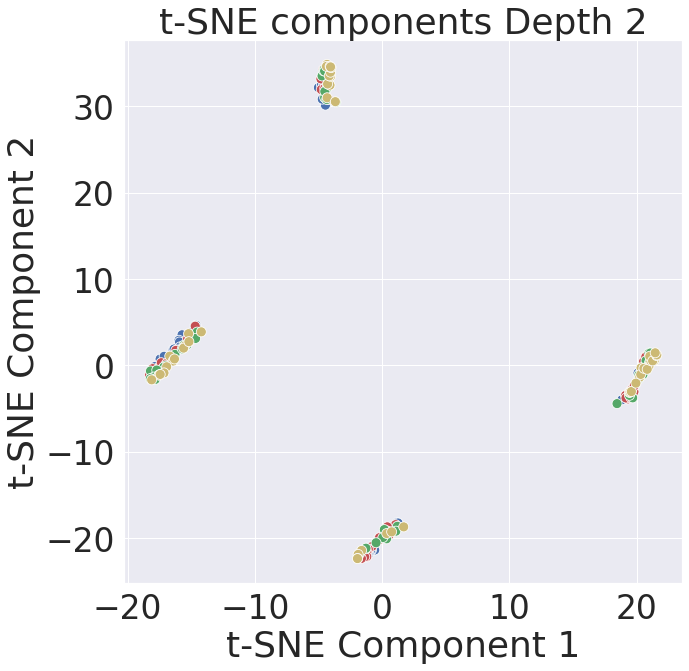}}
\subfloat[]{\includegraphics[width=0.3\textwidth]{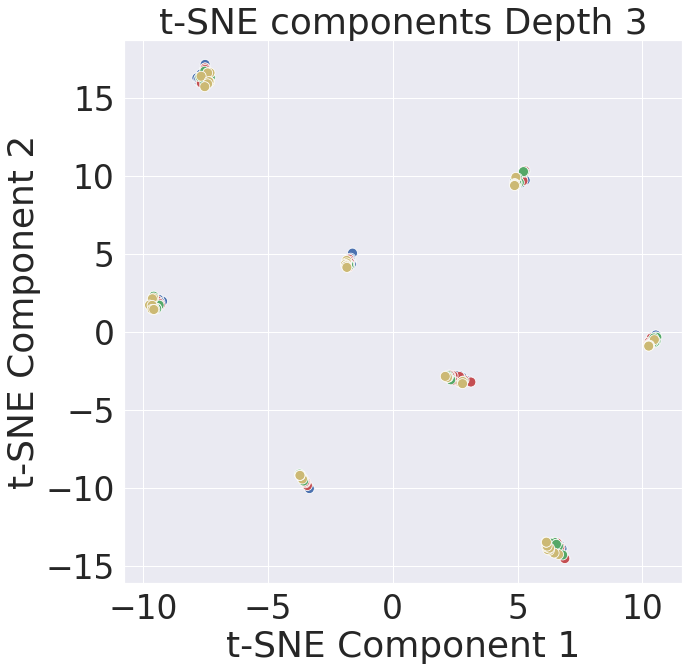}}
\newline
\subfloat[]{\includegraphics[width=0.3\textwidth]{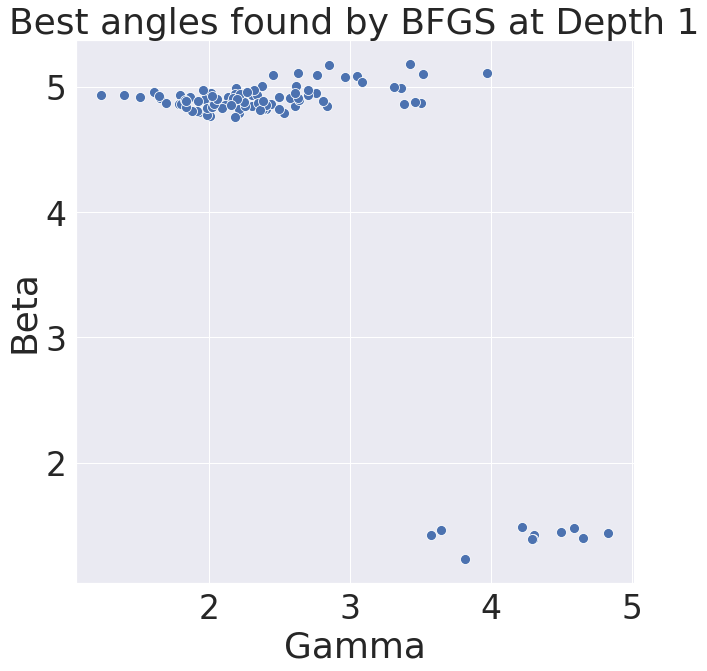}}
\subfloat[]{\includegraphics[width=0.3\textwidth]{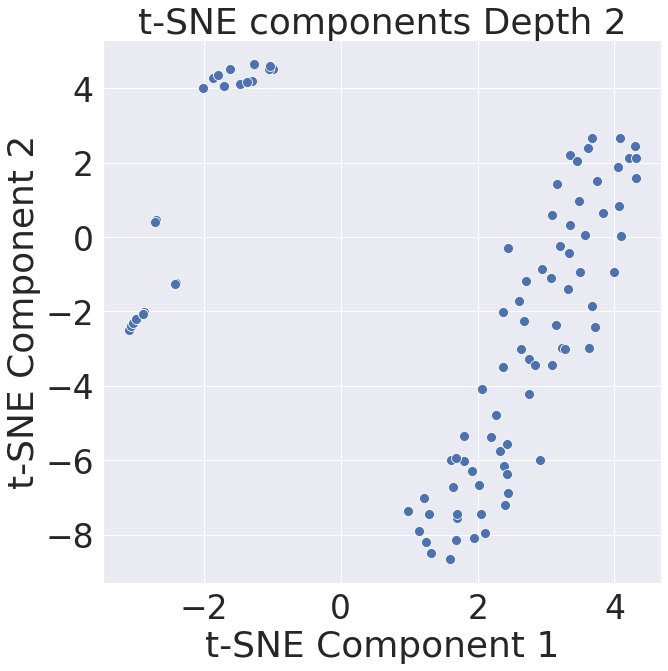}}
\subfloat[]{\includegraphics[width=0.3\textwidth]{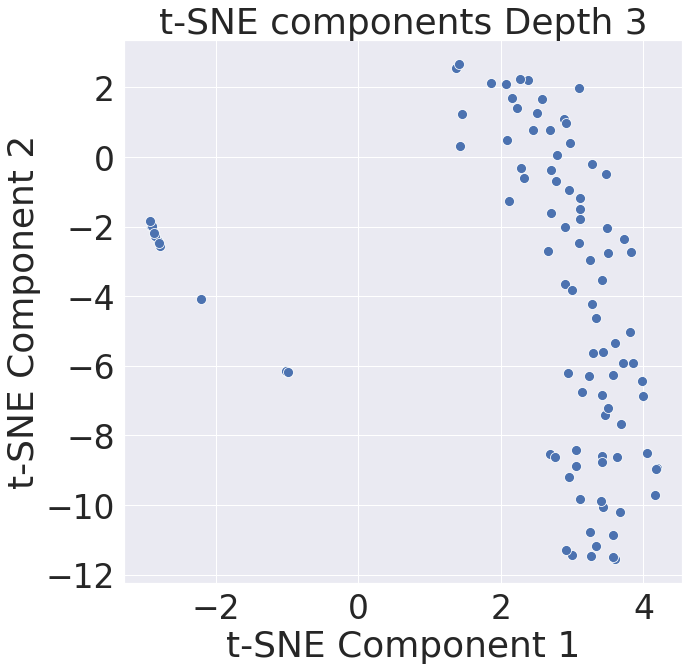}}
\caption{
\small 2D angles visualization $\gamma_i, \beta_i$ for each depth. Plots a), b) and c) concern MaxCut problems while the others refer to the dense QUBO matrices. For $p=2,3$, t-SNE is applied for projecting the angle values to 2D. Different areas of concentration are revealed again. We use different colors for differentiating the probability of edge creation of the Erd\H{o}s-R\'{e}nyi graphs, showing no correlation with clusters.}
\label{tsne}
\end{figure*} 

\section{Clustering-based (unsupervised) learning for angles}
\label{clustering}

As the EDA highlights a clustering effect, we propose different clustering approaches that use different data for angle recommendations. Namely, we describe first using the angle values directly for building clusters serving as angles to try. Then, we switch to using instance-related features. Finally, for the unweighted case, we use graph auto-encoders whose outputs can be used for clustering instead of computing graph features. In the following, we detail each clustering approach for flexible angle recommendation. 

\subsection{Identifying clusters of angles or problem instances}

We first considered clustering using angle values. Given a database of optimal angles for $Q$ problem instances $ \{ I_1, \cdots, I_Q \}, \{ (\gamma^{\ast} , \beta^{\ast})_1, \cdots,(\gamma^{\ast} , \beta^{\ast})_Q \}$, this can be seen as computing or selecting a good set of angle values the database to apply on new instances. In this case, we do not use the problem instances during clustering. Given a user-specified number of angles to be tested $K$, this set of angle values is then applied to new QAOA circuits. To specify them, we can use a clustering algorithm on the database $\{ (\gamma^{\ast} , \beta^{\ast})_1, \cdots, (\gamma^{\ast} , \beta^{\ast})_Q \}$. For instance, K-means \cite{km} will output centroids to use directly as angle recommendations for QAOA on new instances. The K-means algorithm aims to partition a set of $n$ data points $x_i$ into $K$ disjoint clusters $C$, characterized by the mean/centroid of the points within a cluster, denoted $\mu_j$. The partition $P = \{P_1, P_2, \ldots, P_K\}$ ($\forall i\neq j \in [1..K], P_i \neq \emptyset, P_i \cap P_j=\emptyset, \cup_i P_i = \{x_i\}_{i=1}^n$) is chosen by minimizing the within-cluster sum of squares, i.e., $\operatorname{arg\,min}_{P}\sum_{i=1}^K\sum_{x\in P_i}||x - \mu_i||^2$, where the centroid $\mu_i = |P_i|^{-1}\sum_{x \in P_i}x$. The algorithm iteratively updates the centroids by assigning each data point to its nearest centroid and computing the mean, until convergence.
\par
To incorporate knowledge from instances when recommending angles, we change the data fed to the clustering algorithm. We distinguish computing instance features from learning an embedding, that is a user-defined $F-$dimensional representation or encoding of the instances as data. We denote an encoding of an instance $I_t$ as $f(I_t)$.
The angle recommendation framework using a clustering algorithm for such instance representation is presented in Alg.~{\ref{algorecommend}}. First, clusters are learned from the encodings extracted from training data. Then, we find the instances in the database that are the closest in distance to the clusters, and their corresponding optimal angles \footnote{Since the clustering algorithm outputs encodings that do not contain QAOA angle information, we use the QAOA angles of the closest training instances to the clusters.}. The latter are then used for QAOA circuits on new instances, from which we keep the best QAOA output.  

\begin{algorithm}[!ht]
   \caption{$K$-angle recommendation framework for QAOA.}
   \label{algorecommend}
\begin{algorithmic}
   \STATE {\bfseries Input:} Clustering algorithm, number of clusters $K$, 
   \STATE {\bfseries Training Data:}  $\{I_1, \cdots, I_Q \};\{ (\gamma^{\ast} , \beta^{\ast})_1, \cdots, (\gamma^{\ast} , \beta^{\ast})_Q \}$,
   \STATE {\bfseries Testing Data:}  $\{I^{'}_{1}, \cdots, I^{'}_{R} \} $,
   \item[]
   \STATE Initialize $anglesToRecommend = [], encodings = []$
   \FOR{$t=1$ {\bfseries to} $Q$}
   \STATE Compute $f(I_t)$ and append to $encodings$.
   \ENDFOR
   \STATE Apply Clustering algorithm on $encodings$
   \FOR{$c=1$ {\bfseries to} $K$}
   \STATE Get encoding of cluster $f^c$ from the clustering algorithm
   \STATE Get closest point in $encodings$ to $f^c$ and extract index $i_c$
   \STATE Append $(\gamma^{\ast} , \beta^{\ast})_{i_c}$ to $anglesToRecommend$
   \ENDFOR
   \item[]
   \FOR{$t=1$ {\bfseries to} $R$}
     \STATE $bestOutput_t = INF$
     \FOR{$c=1$ {\bfseries to} $K$}
     \STATE Apply QAOA on $I^{'}_t$ with $anglesToRecommend[c]$
     \IF{$E_{anglesToRecommend[c]}(C_{I^{'}_{t}}) < bestOutput_t$}
     \STATE $bestOutput_t = E_{anglesToRecommend[c]} (C_{I^{'}_{t}})$
     \ENDIF
     \ENDFOR
   \ENDFOR
\end{algorithmic}
\end{algorithm}

\subsection{Instance encodings}

In this work, we show two main approaches to encoding the instances for clustering.
First, we computed a set of features following \cite{dunning, Moussa2021}. Such features were used in~\cite{dunning} to decide among classical heuristics to solve MaxCut and QUBO problems. Inspired by~\cite{dunning}, the features were also used for choosing when to apply QAOA against a classical approximation algorithm~\cite{Moussa2021}.
 \emph{For Erd\H{o}s-R\'{e}nyi graphs, we took the graph density, the logarithm of the number of nodes and edges, the logarithm of the first and second-largest eigenvalues of the Laplacian matrix normalized by the average node degree and the logarithm of the ratio of the two largest eigenvalues. For QUBOs, we reduced them to the MaxCut formulation and used the logarithm of the number of nodes, and the weighted Laplacian matrix eigenvalues-based features.}
\par 
We also show how to use graph embeddings using Graph Neural Networks (GNNs) \cite{gnnreview}, avoiding the need for the user to have to compute the features. We employ the Variational Graph Auto-Encoders (VGAE) \cite{kipf2016variational}. This technique only works on unweighted graphs by its design principle. Consequently, we only applied it to the MaxCut instances later in this work. a VGAE learns latent embeddings $\mathbf{Z} \in \mathbb{R}^{N\times F}$ where $F$ is the dimension of the latent variables and $N$ the number of nodes. Given the adjacency matrix $A$ and nodes feature vector $X$, the model outputs the parameters of a Gaussian distribution $\mu, \sigma$ for the latent representation generation. We feed to the model the Erd\H{o}s-R\'{e}nyi graphs, and we add as node features the degree of the nodes. Once learning is completed, we compute the embeddings by a common average readout operation \cite{gnnreview, wang2019dgl}. The latter operation can be defined as averaging the node embeddings for a graph with vertex set $\mathcal{V}$ $\frac{1}{|\mathcal{V}|}\sum_{n\in \mathcal{V}}Z_n$. This allows having a fixed dimension $F$ for the encoding to be used by a clustering algorithm.
\par 
Having defined different strategies for clustering, we apply them to the data we generated and compare their performances. In the following section, we present our results obtained by taking a Machine Learning approach, starting from a simple baseline and cross-validating each method. 

\subsection{Results}

In this section, we apply the above-mentioned proposed strategies to the generated data where EDA revealed different areas of concentration. As the first baseline for angle setting strategy, we experiment with simple aggregation of angle values (median and average). Then we follow this up by K-means by varying the number of clusters from $3$ to $10$ as the underlying clustering algorithm. Finally, we change the K-means data to cluster based on instance encodings instead of angle values. We computed first a set of graph features that were used in a previous study \cite{qalgoselection}. Then we investigate graph autoencoders to learn the encodings of the Maxcut instances. We cross-validate each method using 5-fold cross-validation where we report the ratios $ \frac{(C_{opt} - E_{\gamma,\beta} (C))} {(C_{opt}- E^{cluster}_{\gamma,\beta} (C))} $ on test instances. A value higher than $1$ would mean that the average cost yielded by clustering has improved over the one found by optimization. We also consider the case where one trains on smaller instances to apply to the bigger ones.

\subsubsection{From angle values}

As simple baseline, we compute the average and the median of the optimal angles from the database $\{ (\gamma^{\ast} , \beta^{\ast})_1, \cdots,(\gamma^{\ast} , \beta^{\ast})_Q \}$. From depth-aggregated results, averaging the angle values yielded a median ratio of $0.524$ for MaxCut and $0.672$ for QUBOs, while taking the median values increased it to respectively $0.950$ and $0.941$. This can be explained by the fact that the median value is statistically more robust than the mean when handling data sets with large variability.
\par
\begin{figure*}[!ht]
\centering
\subfloat[]{\includegraphics[width=0.48\textwidth]{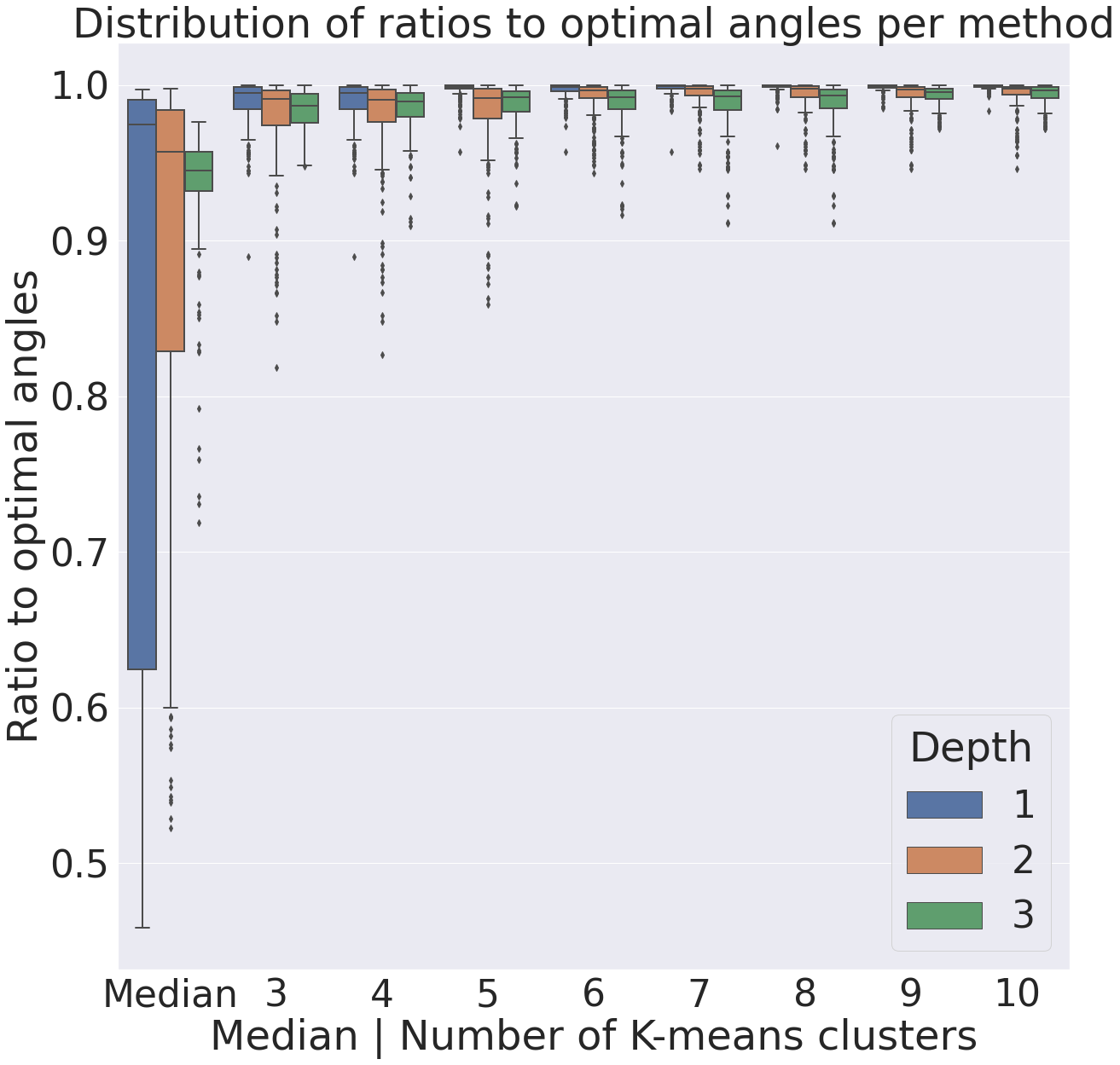}}
\subfloat[]{\includegraphics[width=0.48\textwidth]{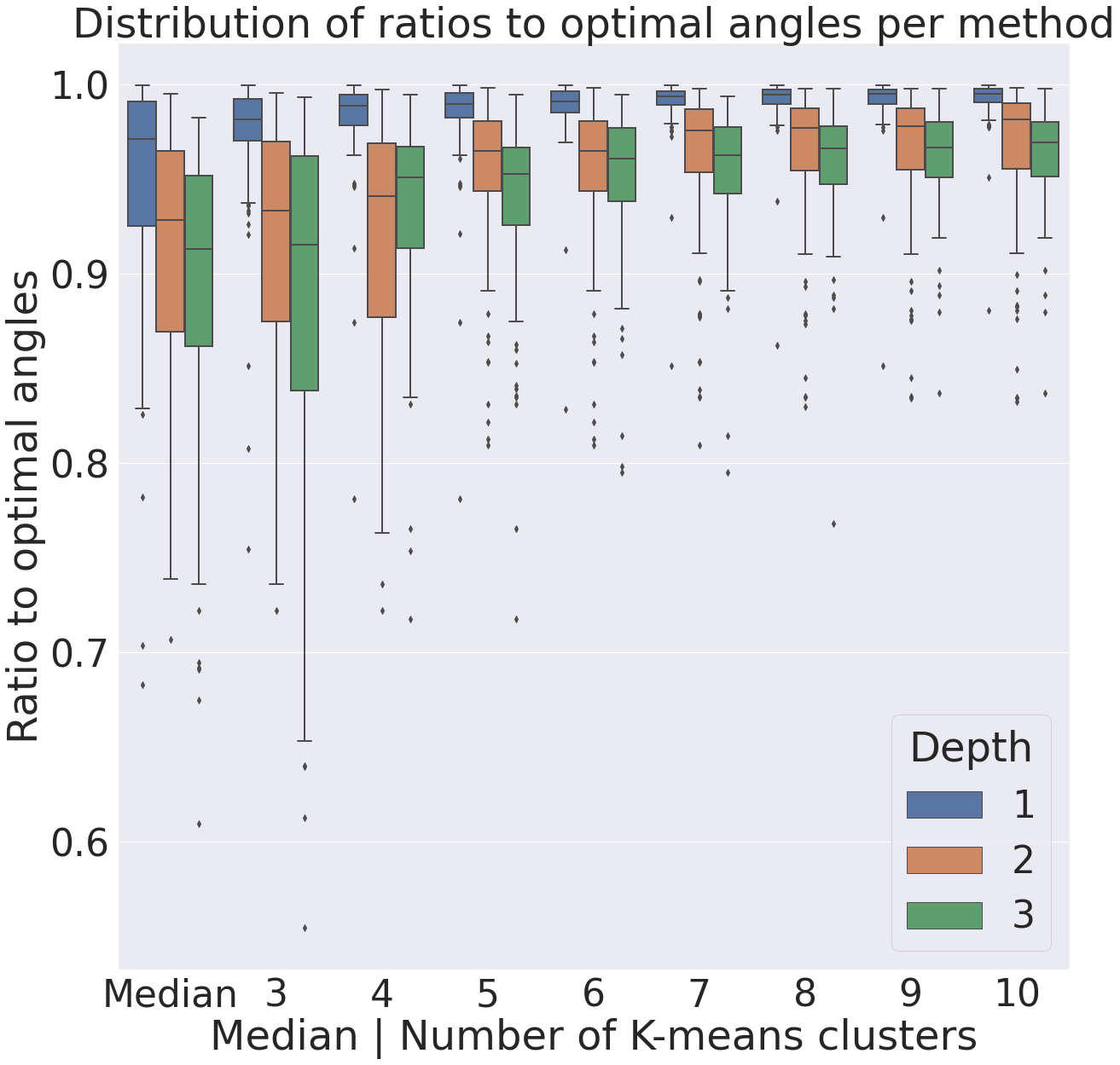}}
\caption{
\small Boxplot visualization of ratios to optimal angles' expectation value per clustering method and depth on MaxCut (a)  and dense QUBOs (b), when using the angle values. We show also the boxplots when computing the median angle values, yielding a median ratio of $0.524307$ for MaxCut and $0.671572$ for QUBOs. The median ratios are respectively $0.990579$, $0.995610$ and $0.998293$ for $3$, $5$ and $10$ clusters. For QUBOs, we get $0.956099$, $0.970842$, and $0.984787$ taking the same number of clusters. With reference to the optimal angles' expectation value, this corresponds on average to a less than $1-2\%$ reduction in performances when using $10$ clusters. For MaxCut, we had to use the closest data point in the dataset to the cluster, as it results in better performances. For instance, with $3$ clusters at $p=3$, the median ratio was $0.618275$.}
\label{clusterstats}
\end{figure*}

As expected with K-means, increasing the number of clusters yielded better median ratios. With $K=10$, the median ratios are $0.998$ and $0.985$ on each dataset, a less than $1-2\%$ reduction in performances w.r.t. the optimal angles. Fig.~\ref{clusterstats} shows the improvement with increased number of clusters. We observe also that with increased depth, median ratio performances are reduced. We conjecture that, when the dimension of the parameter space increases, more clusters are naturally needed to ensure a sensible recommendation. 
\par
Also, such a deterioration of performance w.r.t. circuit depth is more substantial on the QUBO instances than on the MaxCut ones, which can be explained by the clustering patterns in the MaxCut scenario being more significant and regular (Fig.~\ref{tsne}). In addition, this observation suggests that for future work, for dense QUBO instances where the cluster center is not representative for all points pertaining to it, it is more reasonable to take a supervised learning method, which takes the problem instance as input as predicts the optimal angle values.
\par
We also observed that, for the MaxCut problem, the cluster centroid of K-means can be quite distant from the data points when the number of clusters is small and the circuit depth is high. Particularly, this phenomenon deteriorates the median ratio by ca. $30\%$ for $3$ and $4$ clusters with $p=3$. Hence, we decided to take the closest data point to the centroid in each cluster as the recommendation, which solves this issue. For QUBOs, using the cluster centroids directly yields better results.
\par
Overall, increasing the number of angles attempted will improve the quality of the QAOA output. Clearly, the results with less than $4$ clusters present examples where the ratio is low, worsening the median performances. For instance, with $3$ clusters on QUBOs, the median ratio is $0.915$. In the context where the budget of quantum circuit calls is very limited, this could be problematic and call for more robust approaches. To this end, we consider using instance features for clustering.

\subsubsection{From instance encodings}

\begin{figure*}[!ht]
\centering
\subfloat[]{\includegraphics[width=0.48\textwidth]{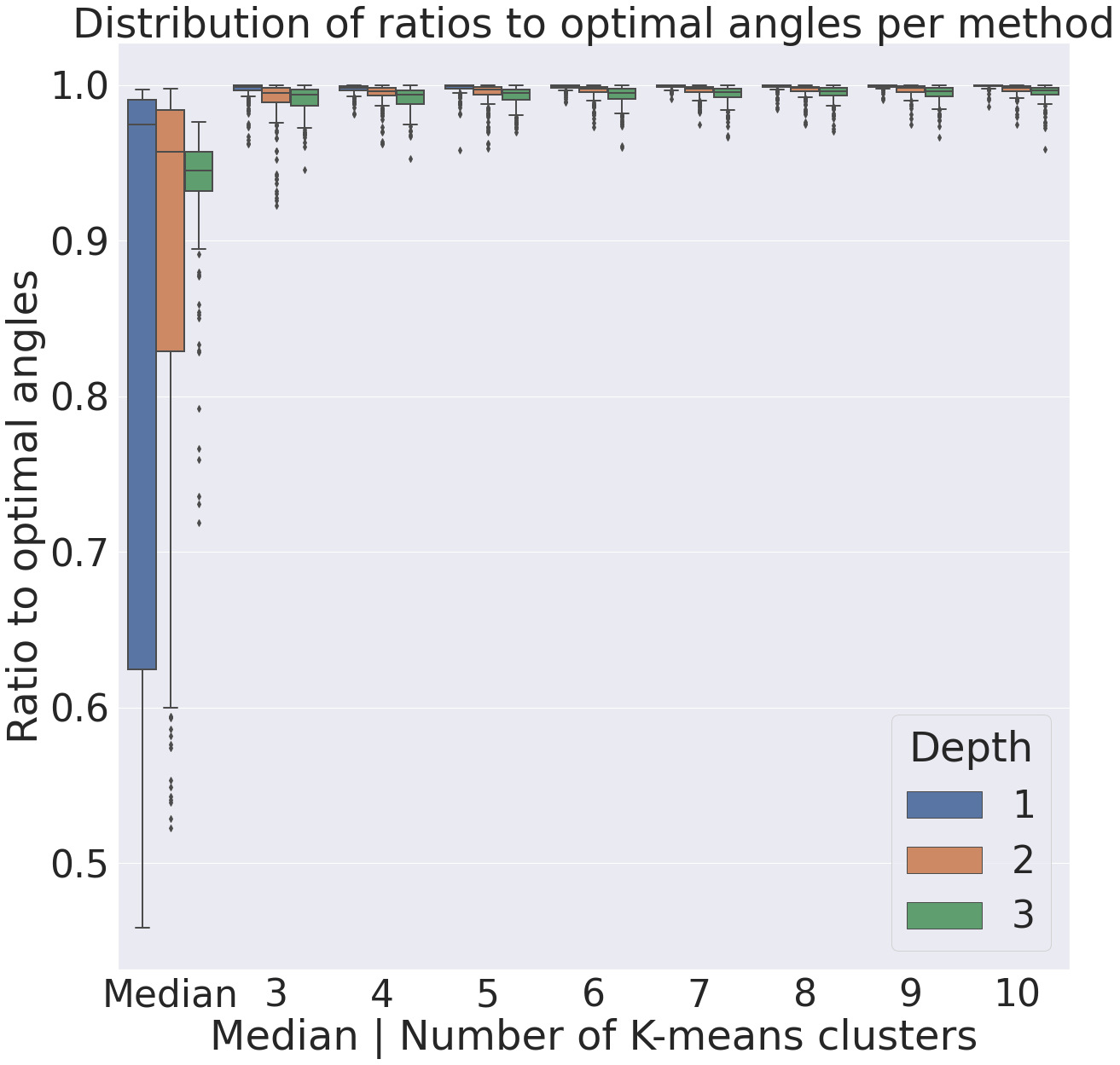}}
\subfloat[]{\includegraphics[width=0.48\textwidth]{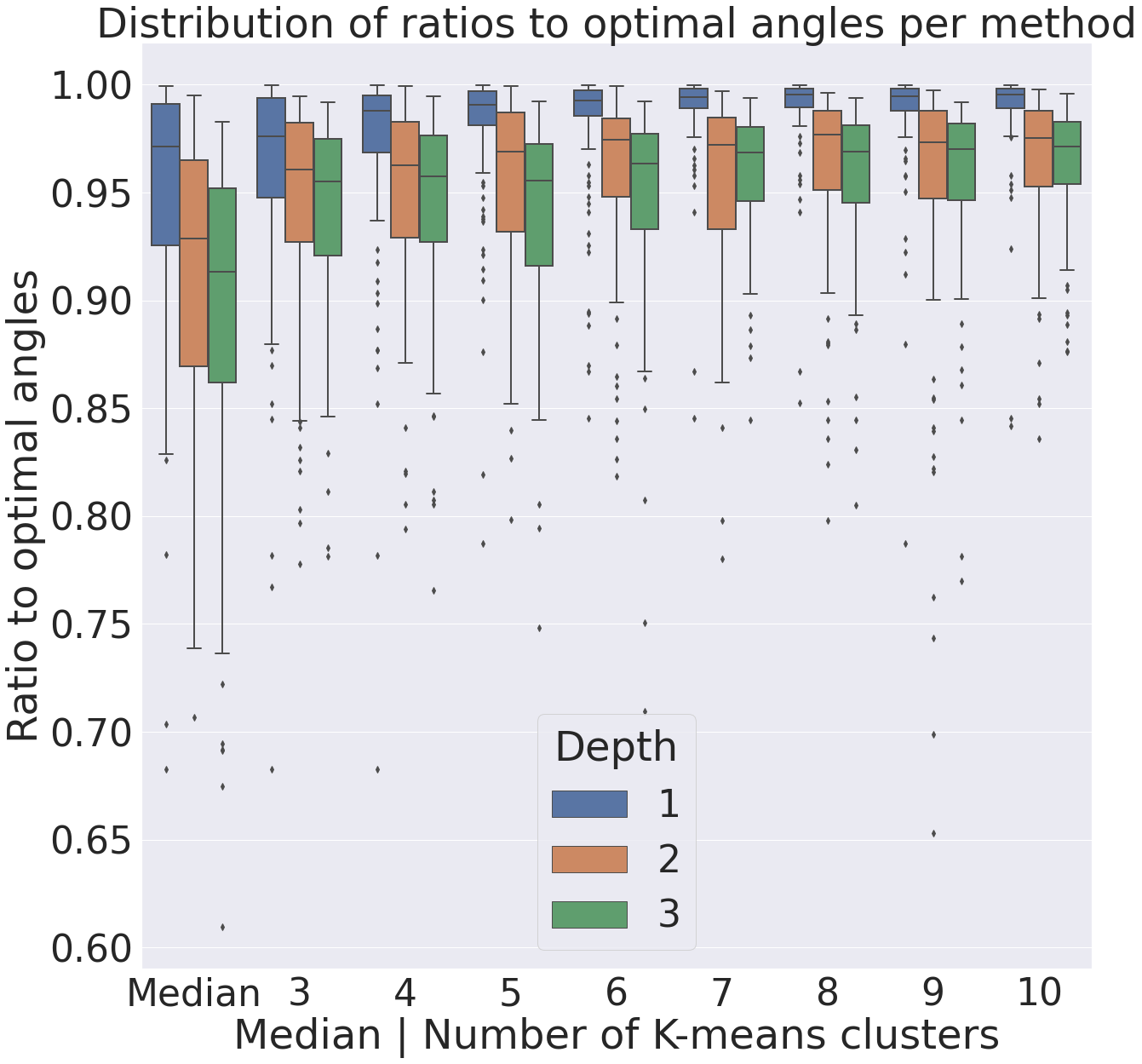}}
\caption{
\small Boxplot plot visualization of ratios to optimal angles' per clustering method and depth on MaxCut (a)  and dense QUBOs (b), when using instance features. For Erd\H{o}s-R\'{e}nyi graphs, K-means yielded ratios $0.996214, 0.996368$ with $3, 4$ clusters and $0.998429$ with $10$. On dense QUBOs, we obtained respective median ratios of $0.963129, 0.971778$ and $0.982964$. With reference to the optimal angles' expectation value, this corresponds on average to a less than $1-2\%$ reduction in performances when using $10$ clusters.}
\label{clusterstatstabular}
\end{figure*}

\begin{figure*}[!ht]
\centering
\subfloat[]{\includegraphics[width=0.48\textwidth]{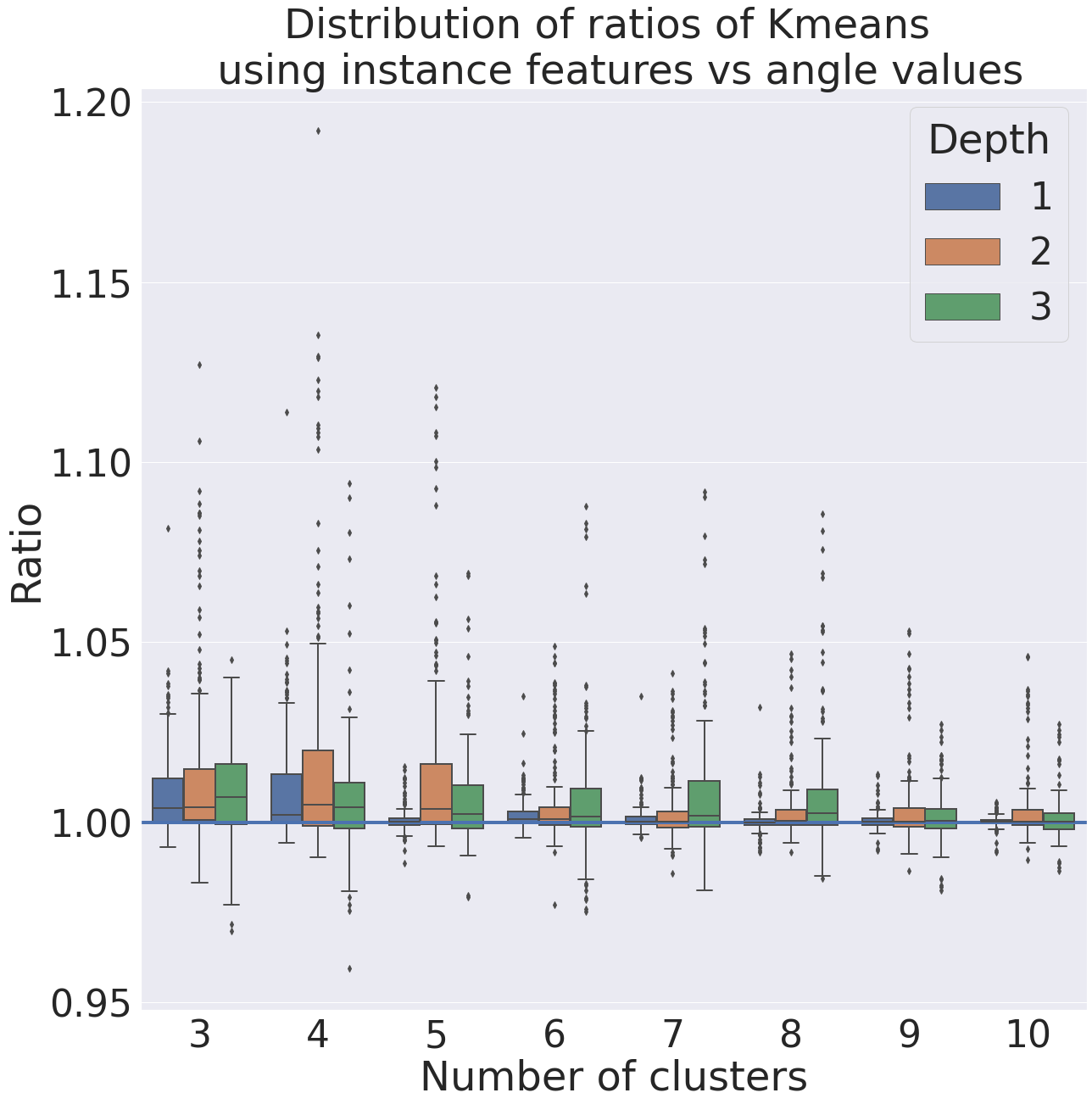}}
\subfloat[]{\includegraphics[width=0.48\textwidth]{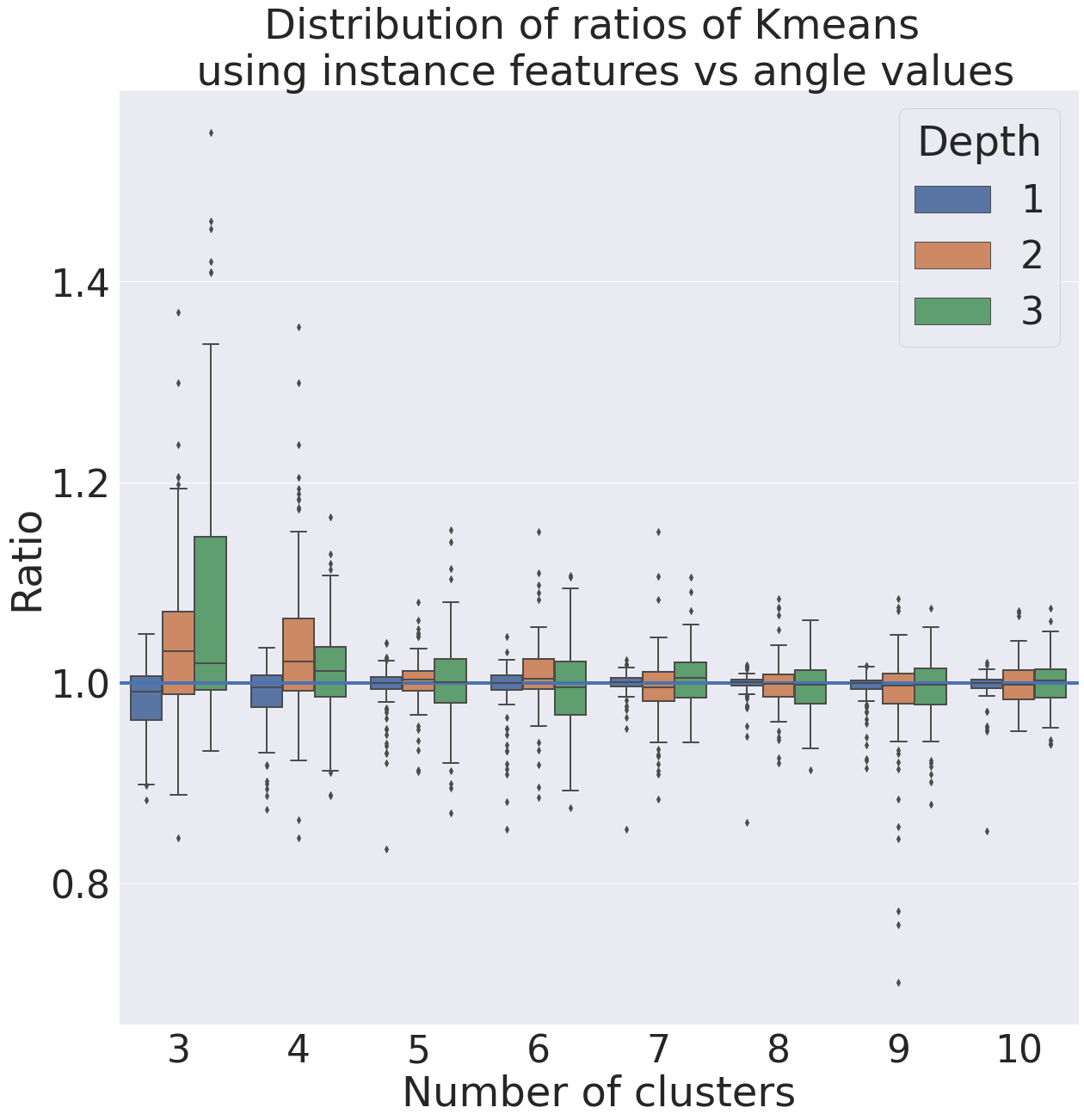}}
\caption{
\small Boxplot of ratios comparing K-means with instance features against angle values on MaxCut (a) and QUBOs (b). A value higher than $1$ (highlighted by a horizontal line) means using instance features results in better QAOA objective. We see an overall improvement with $3,4$ clusters mainly at $p=3$.}
\label{K-meanstabularvsangle}
\end{figure*}

To witness whether using instance features can improve the quality of clustering, we divided the ratios obtained with instance features by the ones using angle values. We show these results in Fig.~\ref{clusterstatstabular} and Fig.~\ref{K-meanstabularvsangle} where we can clearly see better ratios with less than $4$ clusters, and similar results on average otherwise.

\begin{figure}[!ht]
\centering
\includegraphics[width=0.46\textwidth]{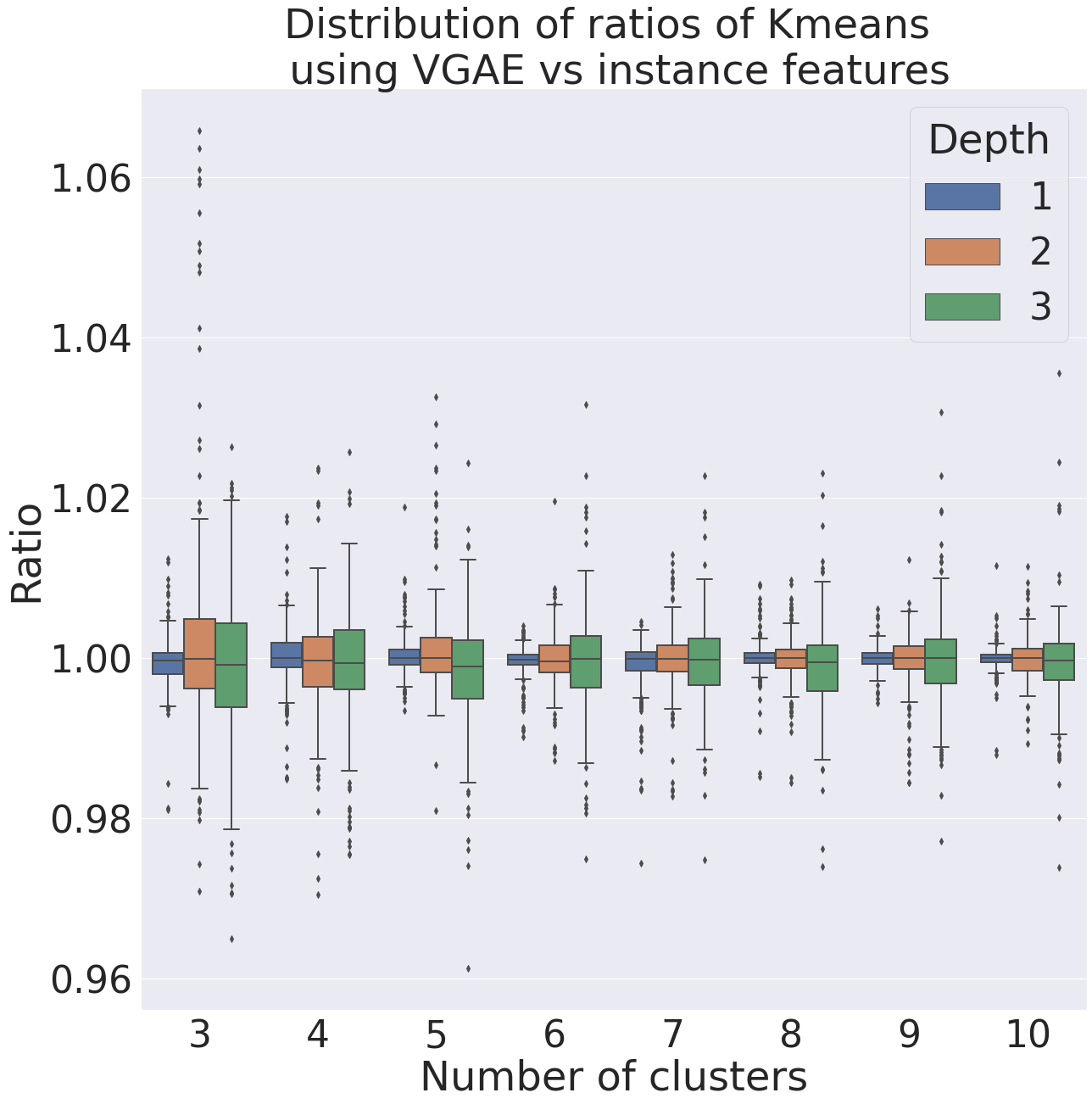}
\caption{
\small Boxplot visualization of ratios on MaxCut obtained using Variational Graph Auto-Encoders compared to using instance features. A value higher than $1$ means using VGAE results in a better QAOA objective. Overall, performances are similar as the ratios are close to $1$ on average.}
\label{gcnvstabular}
\end{figure}

As for learned encodings or embeddings with auto-encoders, the GNN model configuration we use is the same two-layer graph convolutional layer as \cite{kipf2016variational}. Namely, the first one has 32 output-dimension using the ReLU activation function. This is followed by two 16-dimensional output layers for the generation of the latent variables. We train using Adam with a learning rate of $0.01$ for $100$ epochs and batch size set to the dataset size. Our implementation uses the Deep Graph Library (DGL) \cite{wang2019dgl}. The embeddings obtained by averaging are of dimension $F=16$. This allows having a fixed dimension for the encoding as input of the same K-means strategy described above. We observe in Fig.~\ref{gcnvstabular} that the results are similar to the ones obtained using instance features. Yet, in some instances, we see better results. Hence, many clustering results can be combined to improve the performances in ratios canceling each other weaknesses at the cost of trying more angles to find the best ones. As future work, we could also decide which heuristic to use depending on a given test instance by using a ML model.
\par
Finally, our approaches can save numerous circuit calls compared to de novo optimization. The median numbers of circuit calls for the BFGS runs giving the best QAOA angles were $56, 150, 320$ for each depth respectively on MaxCut and $44, 132, 252$ for QUBO, while in the cluster approach, the number of calls is always the cluster size, which is considerably smaller than the cost of BFGS. Instance size does not seem to affect the number of circuit calls by BFGS. In our approaches, we limited circuit calls to $10$ and we do not need multiple restarts.

\subsection{Aggregating results}

\begin{figure*}[!ht]
\centering

\subfloat[]{\includegraphics[width=0.49\textwidth]{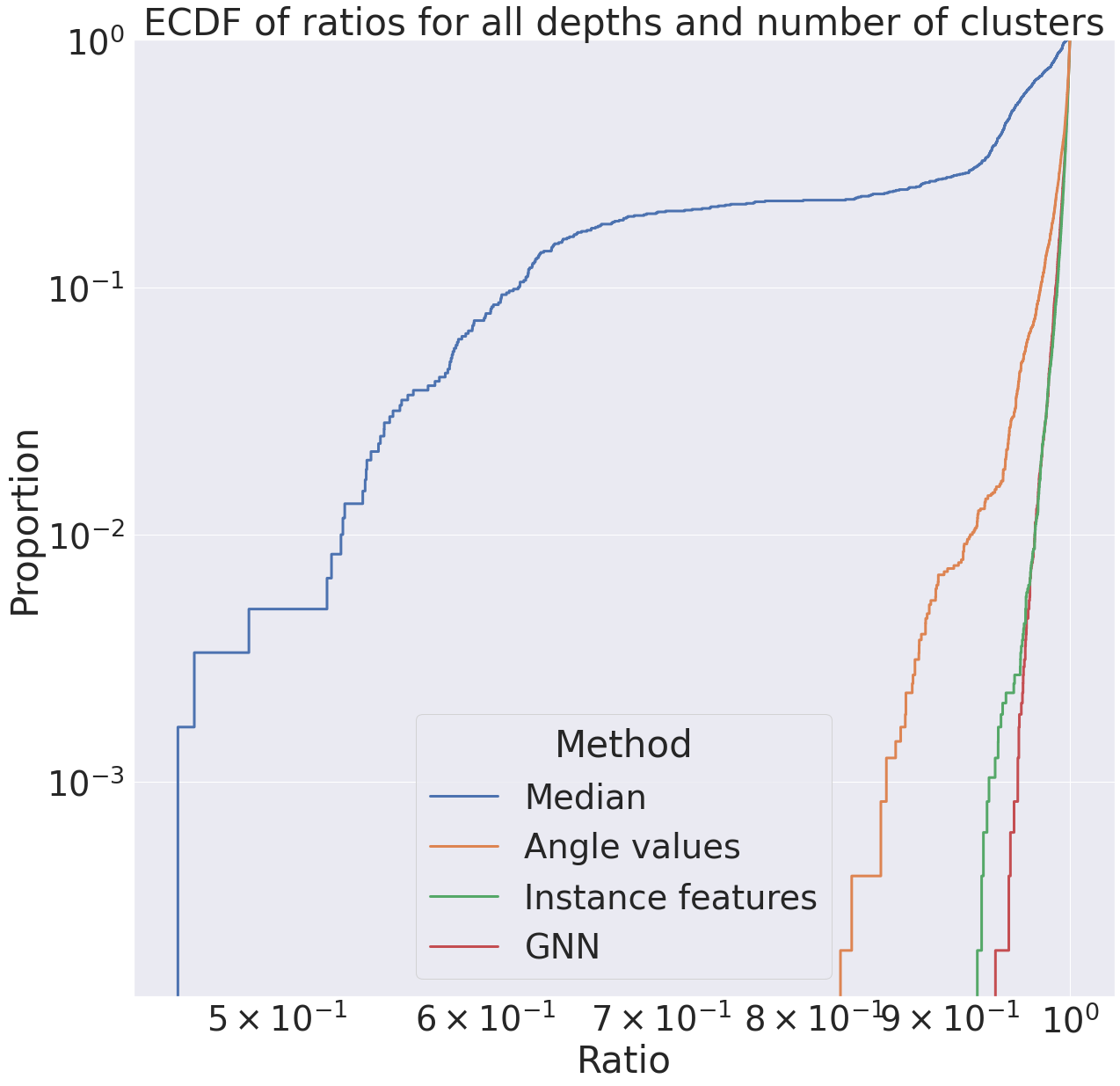}}
\subfloat[]{\includegraphics[width=0.49\textwidth]{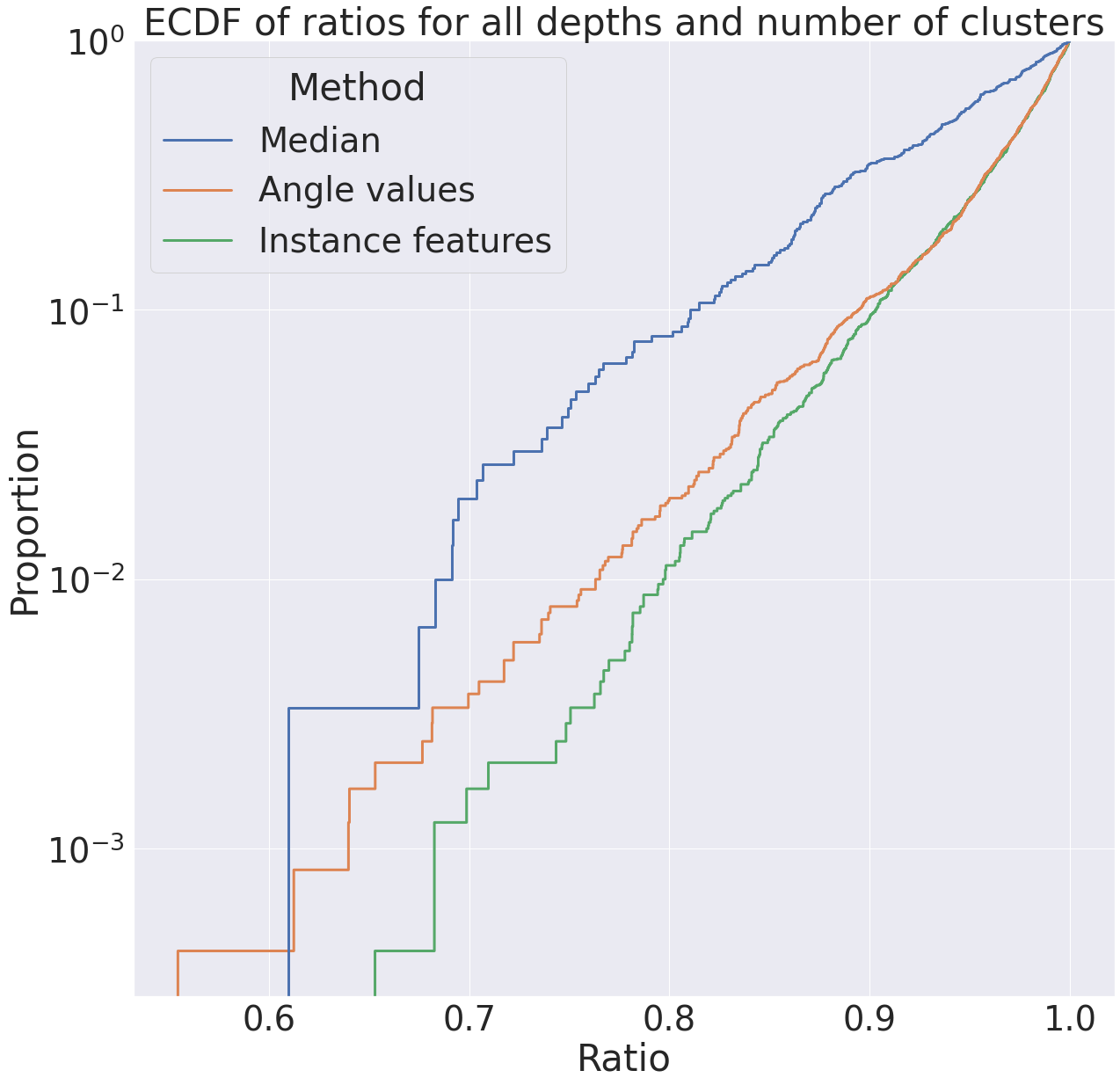}}
\caption{
\small Empirical cumulative distribution functions of ratios to optimal angles' for all depths and number of clusters. A lower curve for an approach means better results when using it aggregating depths and number of clusters. We see for MaxCut (a) and QUBO problems (b) that instance features achieve better results, VGAE being competitive with instance features. When using $3$ clusters, using VGAE on MaxCut instance and instance features for QUBOs lead to better ratios.}
\label{freqs}
\end{figure*}

\begin{figure*}[!ht]
\centering
\subfloat[]{\includegraphics[width=0.49\textwidth]{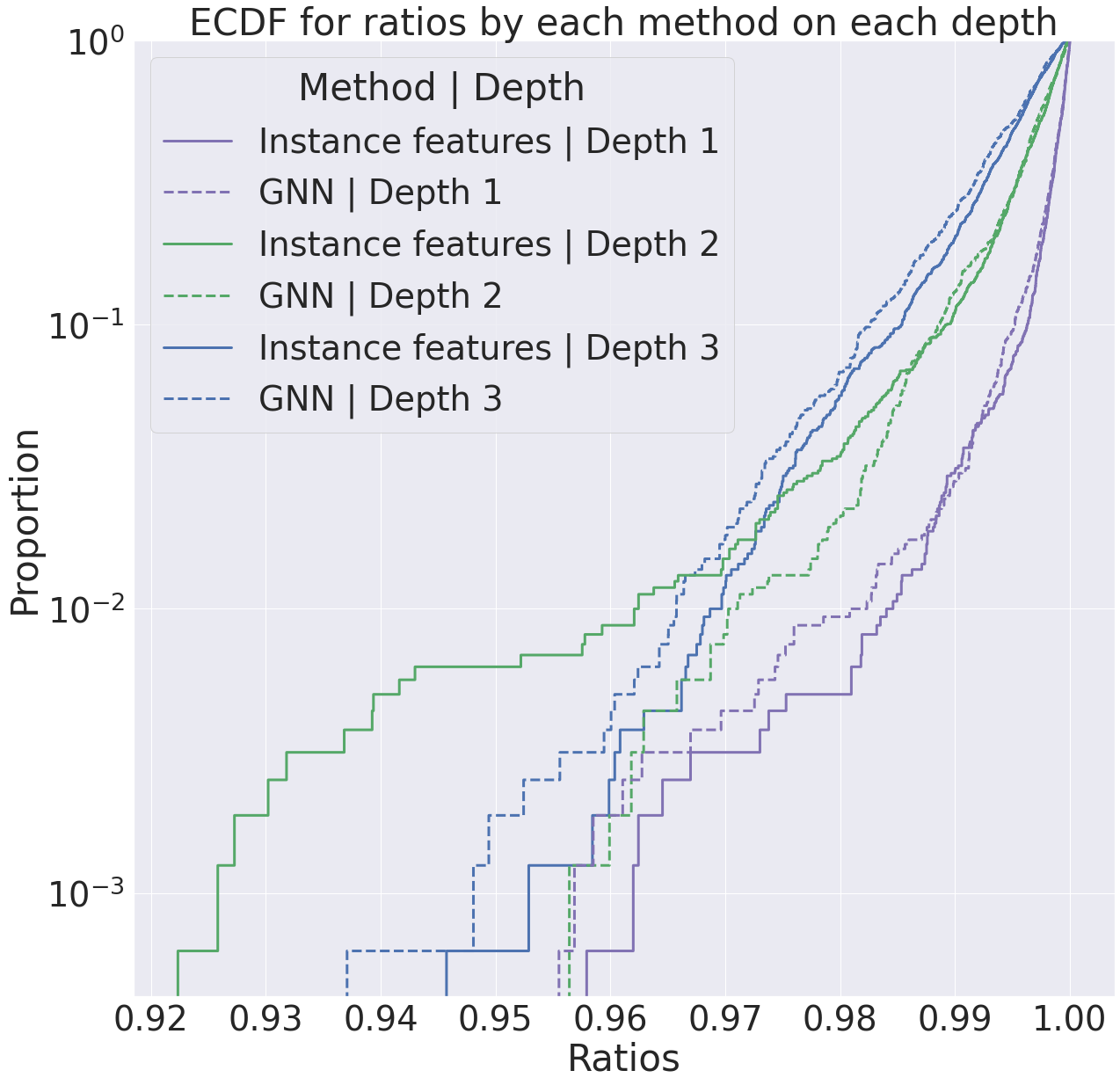}}
\subfloat[]{\includegraphics[width=0.49\textwidth]{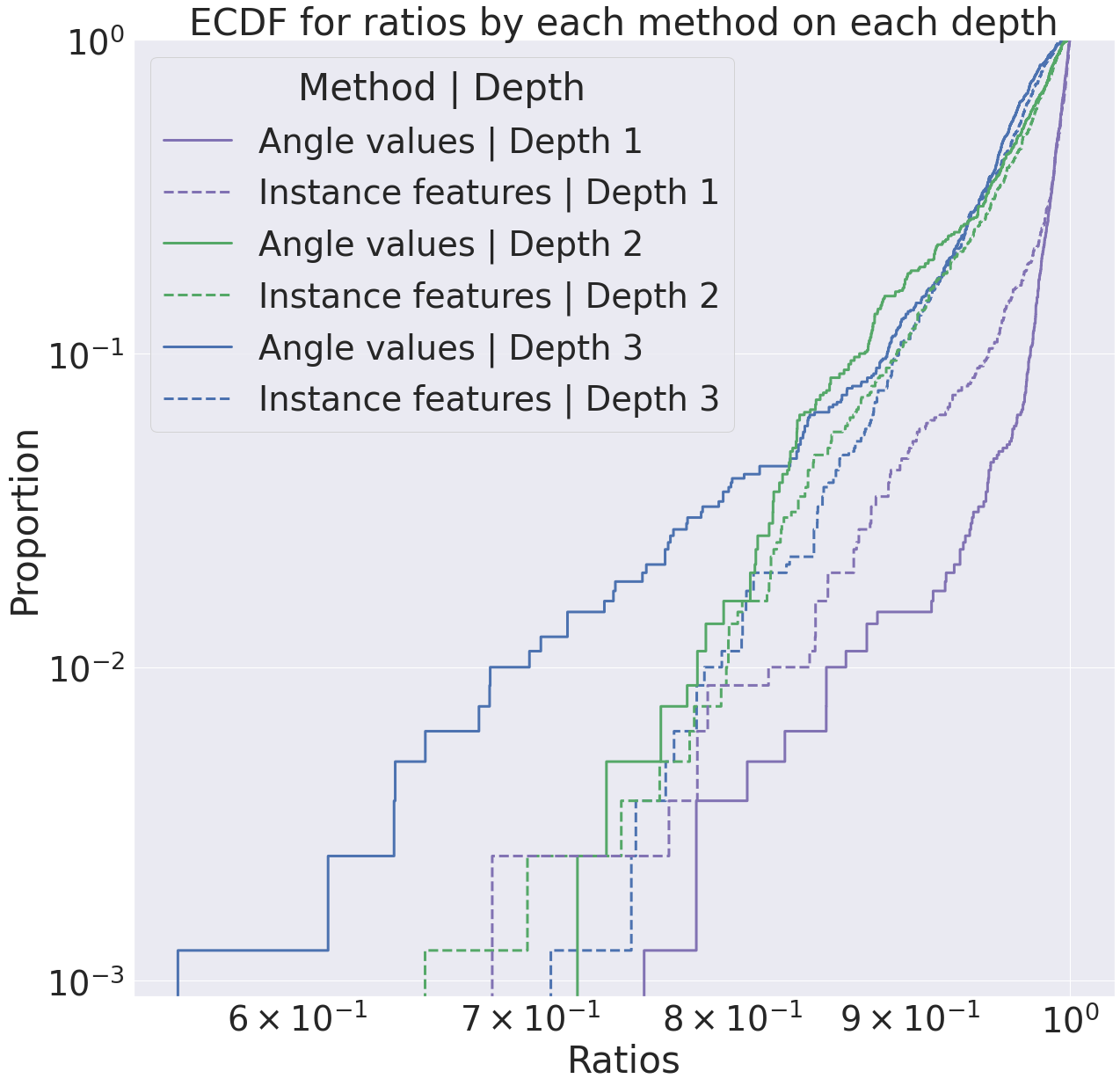}}
\caption{
\small Empirical cumulative distribution functions of ratios to optimal angles per method and depth. The lower the curve, the better the method. In most cases, the curve corresponding to instance features was lower (except for QUBOs (b) at $p=1$, and VGAE's curve was more competitive at $p=2$ for MaxCut (a)). This was also the case when using $3$ clusters.
}
\label{freqsperdepth}
\end{figure*}

Following the presentation of the different clustering approaches, we compare their performances to determine which approach works best. We propose to take the Empirical cumulative distribution functions (ECDF) of the ratios as the performance measure to compare those different approaches. Given a sample $\{r_i\}_{i=1}^R$ of the ratios and a value of interest $t\in[0, 1]$, ECDF is the fraction of the sample points less or equal to $t$: $F(t) = \frac{1}{R} \sum_i \mathds{1}_{[0, r_i]}(t)$, where $\mathbbm{1}{}$ denotes the indicator function, which returns one only if $t\in [0, r_i]$ and zero otherwise. They enable us to aggregate the results of the different numbers of clusters and depths. A better method will have more proportion of higher ratios, resulting in an ECDF curve located more to the right.
From Fig.~\ref{freqs}, we observe that \emph{using instance encodings is more successful in yielding better angles than using the angle values}. This is also witnessed in Fig.~\ref{freqsperdepth} with increased depth and a low number of clusters. Also, VGAE seems to be slightly better than instance features on the MaxCut problems. However, these methods can complement each other, especially as we do not need to increase dataset size. Hence, combining them at the cost of circuit calls becomes an option for running QAOA, as we showcase with RQAOA in the next section.

\subsection{Case when test instances are bigger than training instances}

\begin{figure*}[!ht]
\centering
\subfloat[]{\includegraphics[width=0.48\textwidth]{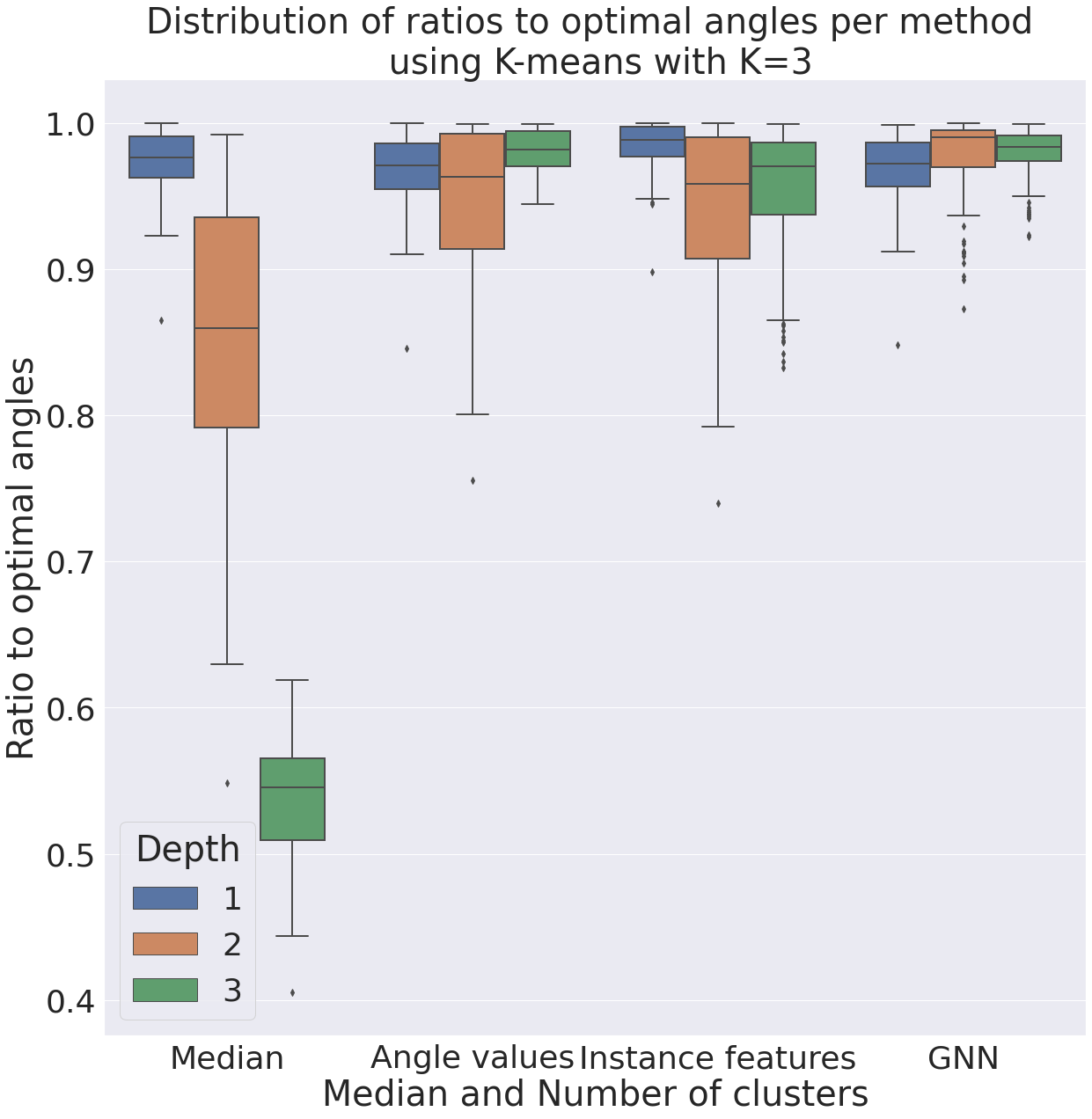}}
\subfloat[]{\includegraphics[width=0.48\textwidth]{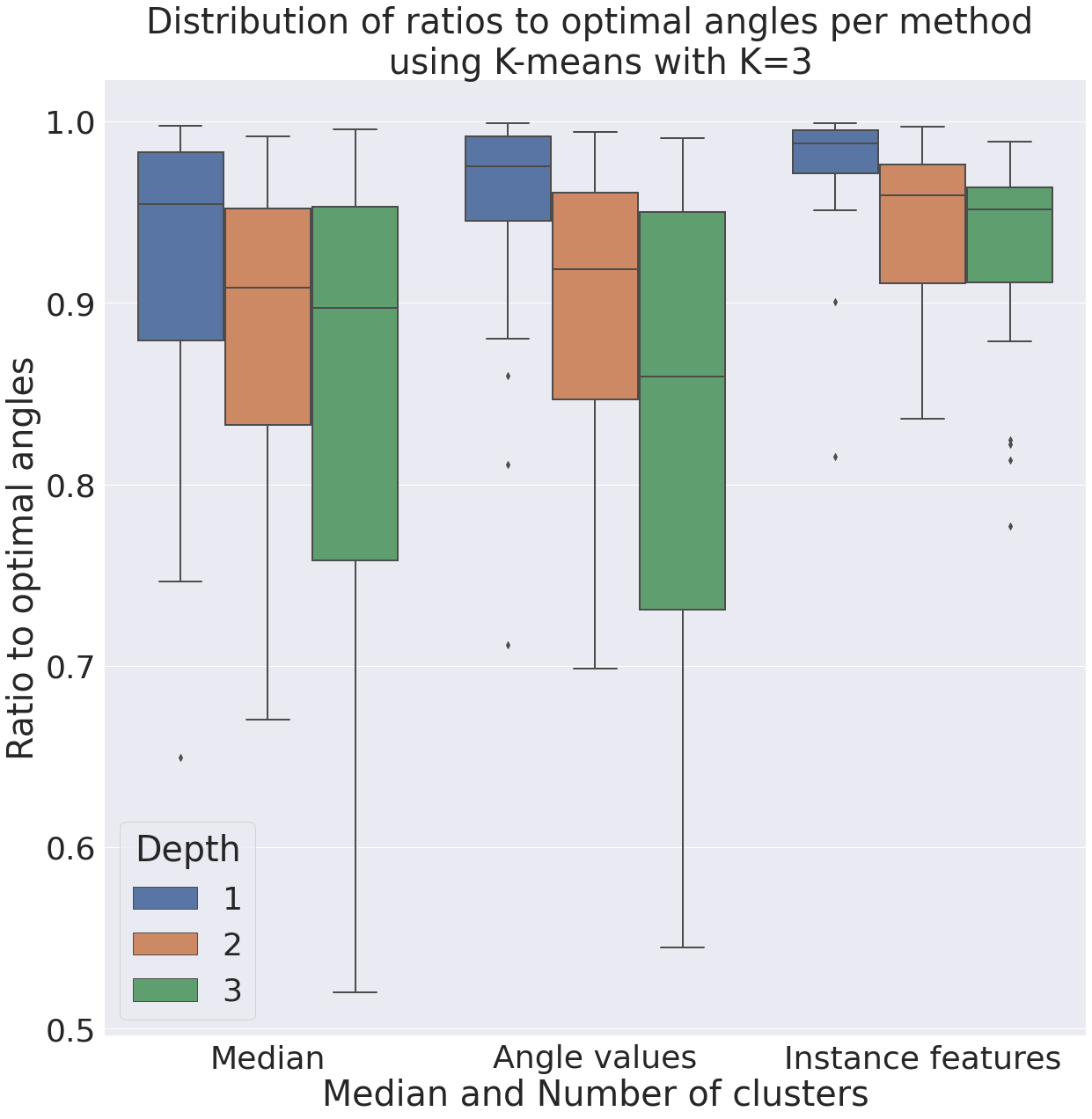}}
\caption{
\small Boxplot of ratios comparing K-means $K=3$ using instance features against angle values on MaxCut (a) and QUBOs (b). The ratios are obtained from $40\%$ of the instances with the highest number of nodes. From depth-aggregated results, on MaxCut, using the median values gives a median ratio of $0.859316$, $0.928519$ with angle values, $0.976959$ with instance features, and $0.981618$ using VGAE. On QUBOs, we obtained respectively $0.926136$ for the median of angle values, $0.936679$ clustering with angle value, and $0.963677$ with instance features.}
\label{smallbigboxplots}
\end{figure*}

\begin{figure*}[!ht]
\centering
\subfloat[]{\includegraphics[width=0.46\textwidth]{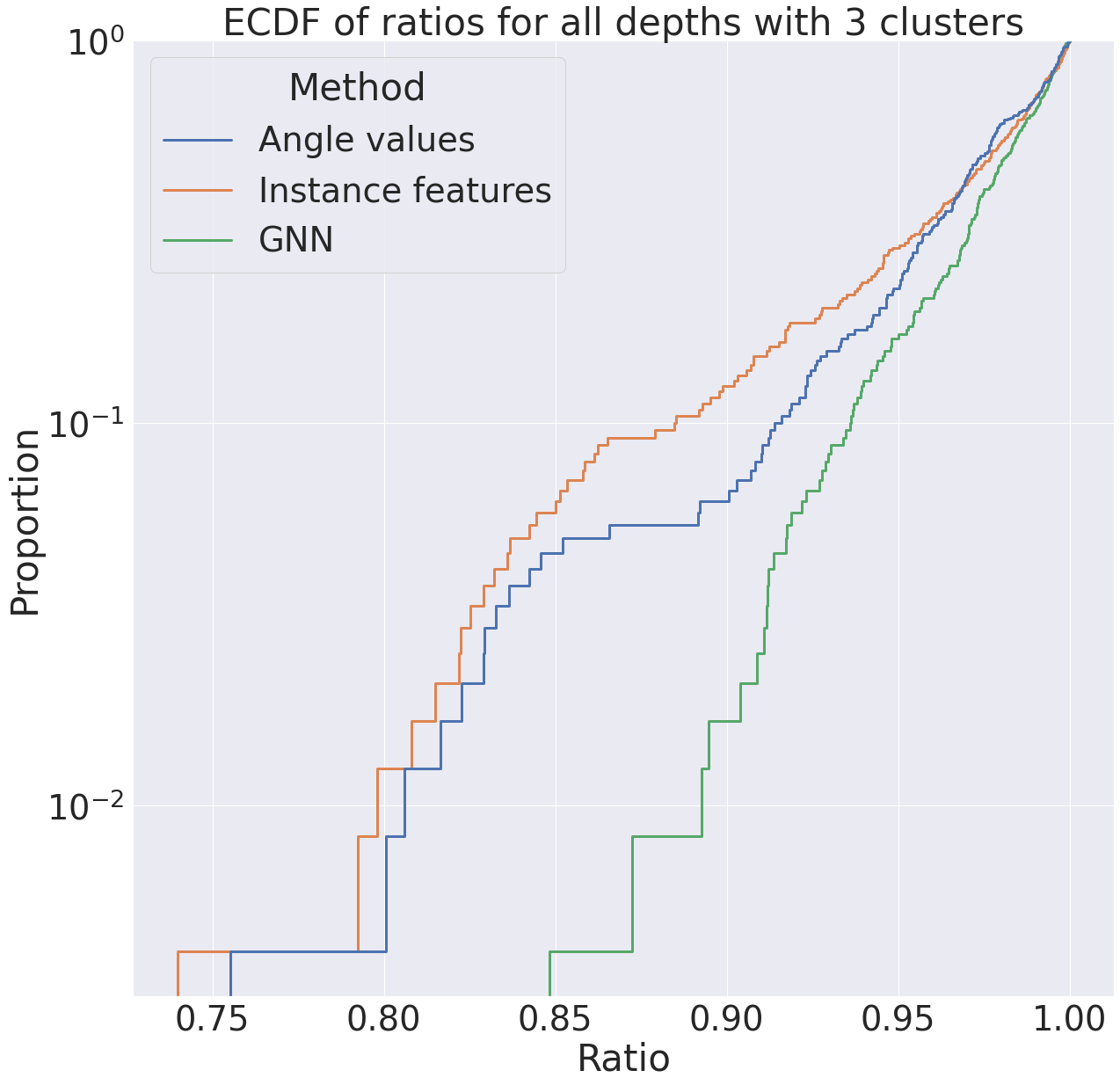}}
\subfloat[]{\includegraphics[width=0.46\textwidth]{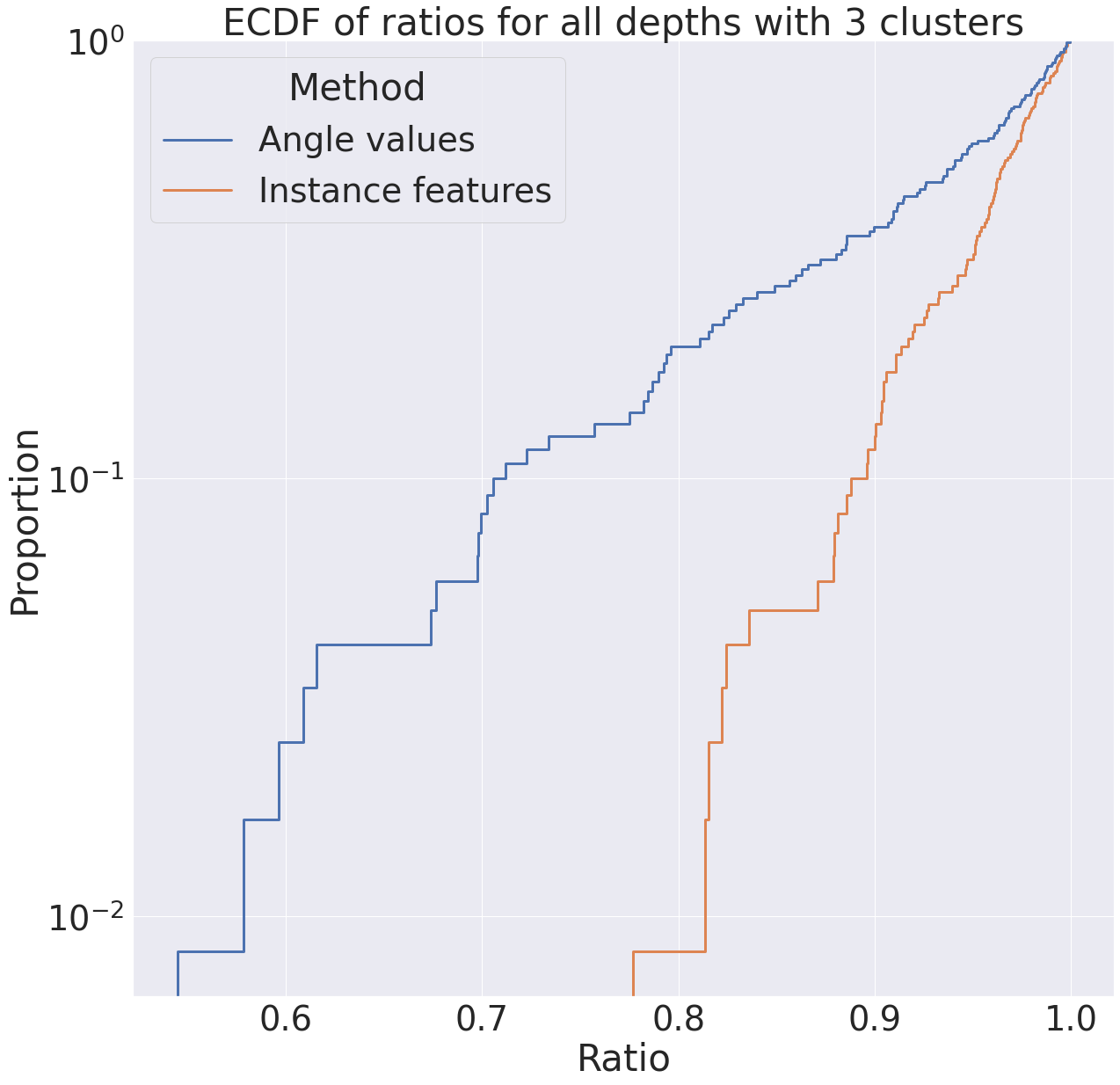}}

\caption{
\small Empirical cumulative distribution functions of ratios to optimal angles'. The ratios are obtained from $40\%$ of the instances with the highest number of nodes. Similar results to Fig.~\protect\ref{freqs} and Fig.~\protect\ref{freqsperdepth} are obtained.
}
\label{smallbigecdf}
\end{figure*}

One important consideration of these methods is to analyze scaling. This is relevant in settings where one is interested in solving larger instances given small ones. In our case, we apply these approaches in the case $K=3$ by a $60-40\%$ train-test split. From Fig.~\ref{smallbigboxplots} and \ref{smallbigecdf}, we find similar conclusions with respectively VGAE on MaxCut and instance features on the QUBO problems yielding better results. Note that we did not use the logarithm of the number of nodes and edges as features when using instance features as the values between training and test are too different.

\section{Demonstration with RQAOA}
\label{resrqaoa}

RQAOA \cite{rqaoa} is a recursive algorithm where, given an Ising problem $\sum_{i,j} w_{ij} Z_i Z_j $, one starts by applying QAOA on the former. the quantum state output $\ket{\mathbf{\gamma},\mathbf{\beta}}$ is then used to compute correlations $ M_{ij} = \bra{\mathbf{\gamma},\mathbf{\beta}} Z_i Z_j \ket{\mathbf{\gamma},\mathbf{\beta}} $. Then, variable elimination is carried out by selecting a pair of variables satisfying $(i_l,j_l) = \operatorname{arg\,max} |M_{ij}|$, and substituting $Z_{j_l}$ with $ \operatorname{sign}(M_{i_l, j_l})Z_{i_l}$ in the Ising formulation. This reduces the number of variables by $1$. We then get a new reduced problem and we reiterate the procedure for a number of user-defined number of iterations. The choice of iteration fixes the size of the final instance which is then solved using a brute-force (or some other classical) approach, and the substitutions are used onto it to obtain a final solution.
\par 
As RQAOA requires optimizing many QAOA instances that iteratively shrink in sizes, we demonstrate the application of our clustering approaches in this context. We do so for the MaxCut problems where we limit the number of iterations to half of the size of the Erd\H{o}s-R\'{e}nyi graphs. We do not consider the dense QUBOs as RQAOA would reduce an original dense graph to non-dense intermediate subproblems not part of the database. As per the number of QAOA parameters attempted per iteration, we limit it to $3$ and apply the three clustering approaches: angle-value, instance features, and VGAE-output based. We do so by using our previous database and training each method on all instances to get $3$ QAOA parameter recommendations. The latter are then used for QAOA on the RQAOA generated instances. 
\par
\begin{figure*}[!ht]
\centering
\subfloat[]{\includegraphics[width=0.46\textwidth,valign=t]{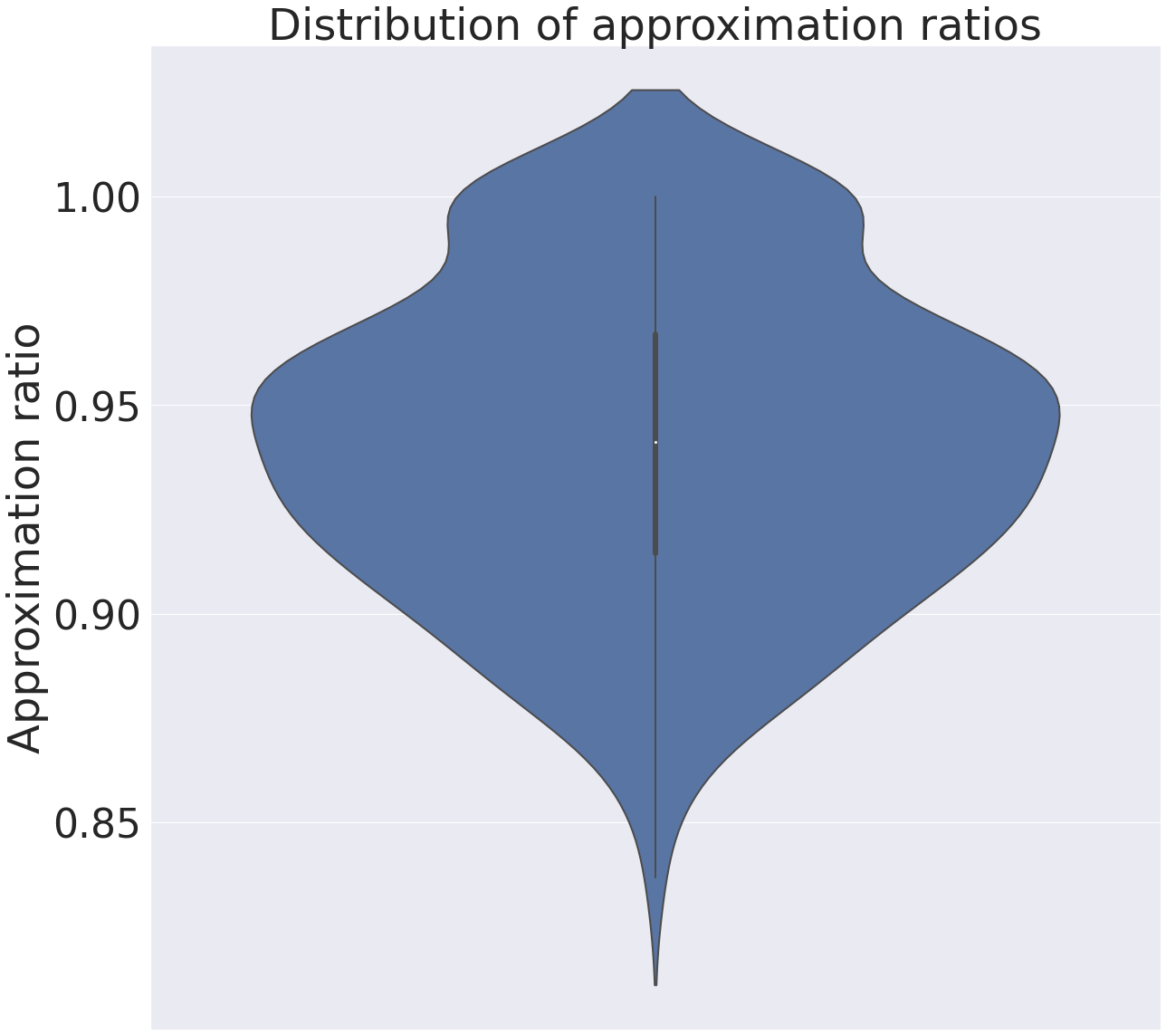}
\vphantom{\includegraphics[width=0.46\textwidth,valign=t]{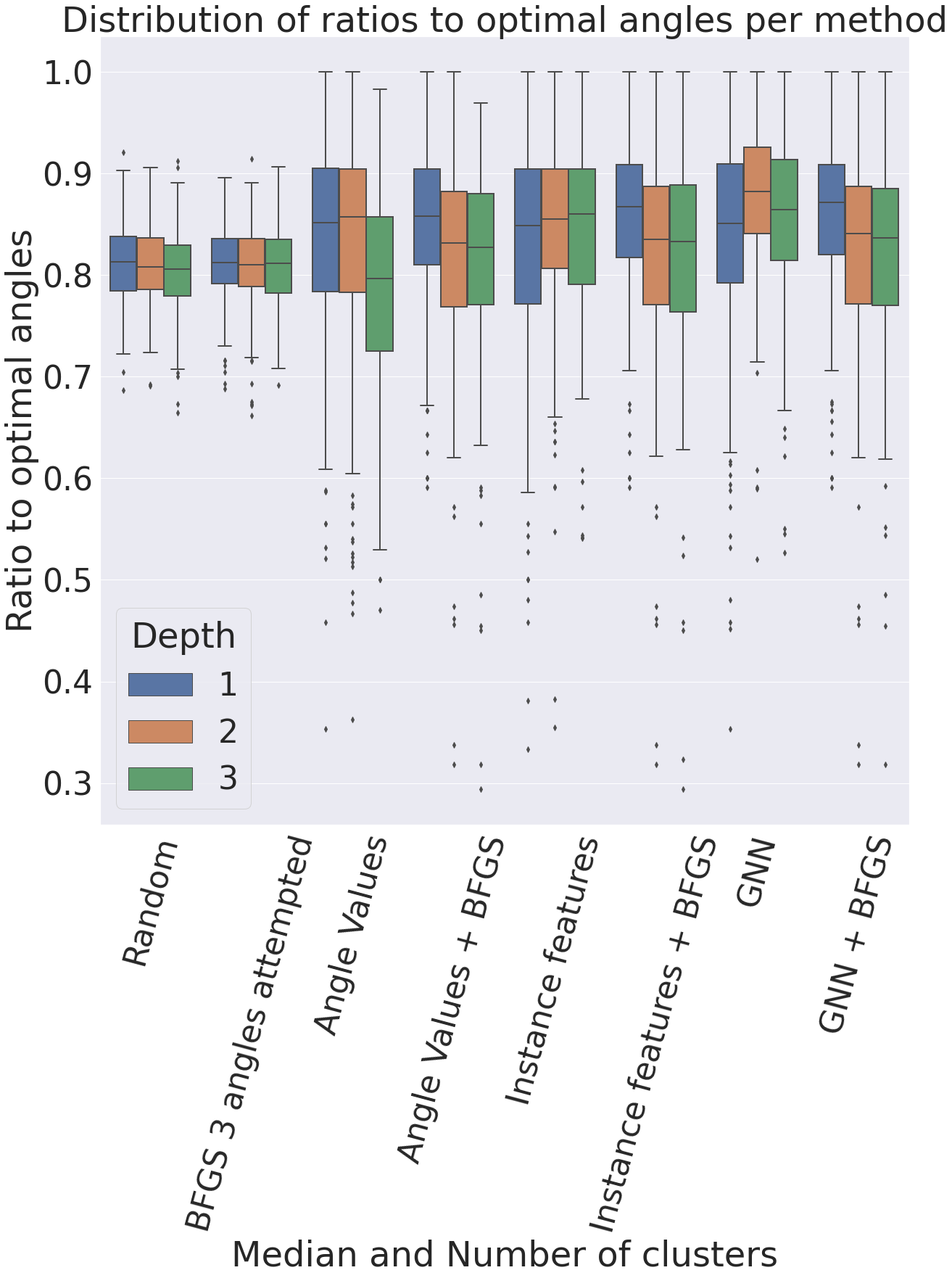}}
}
\subfloat[]{\includegraphics[width=0.46\textwidth,valign=t]{figuresv2/rqaoa/bymethodBFGS.png}}
\caption{
\small The violin plot visualization represents ratios obtained on MaxCut using all three unsupervised approaches, using just $3$ circuit calls per RQAOA iteration, without optimizing further with BFGS. A median ratio of $0.94117$ was obtained. The boxplots represent the MaxCut ratios obtained using each approach. We added using random angles per iteration as a baseline as well as using BFGS (starting with random angles) with a budget of 3 angle values attempted during optimization, and we witness clustering approaches yielded better results. When dividing the ratios of the methods with the ones obtained by adding BFGS, we obtain a median ratio of $1$, meaning we saved many circuit calls for similar results with clustering.}
\label{rqaoaratios}
\end{figure*}

 
Fig. \ref{rqaoaratios} shows that with the three approaches, we obtain a median $0.94117$ approximation ratio with RQAOA. The minimal ratio obtained is $0.8367$ and the optima were found on $33$ instances. When looking at each method independently, we observe that the angle-value clustering performances at $p=3$ are lower than the others. This is due to the fact that we use the K-means clusters directly as it allowed us to find more instances with a ratio of $1$. Graph features and VGAE seem similar in performance, with a small advantage at depth $2$ for VGAE. Looking at the frequencies where the best ratio by instance was obtained, VGAE is more successful. Respectively, each method achieves the best-found ratios over $88$, $118$, and $165$ instances. Finally, we also tried using random angles, by sampling uniformly values in $[0, 2\pi]^p$, and optimizing further the angles from each approach with BFGS up to $100$ iterations maximum. We clearly see better performances with clustering approaches compared to random angles. This is also the case when using BFGS (starting with random angles) limited to $3$ circuit calls when optimizing, the same budget as our clustering-based approaches. Dividing the MaxCut ratios obtained with BFGS with the ones without further optimization yielded a median value of $1$. Hence, the results were similar to the BFGS-optimized approaches, saving many circuit calls.
\par 
To conclude, our unsupervised approaches can be used to run quantum algorithms where QAOA is used as a subroutine. They are then considered as hyper-parameters that can be tweaked to achieve better performances for QAOA-featured algorithms, depending on a user-defined budget definition. In our RQAOA showcase, the maximal depth of QAOA, as well as the number of parameters to try at each iteration, was set to $3$, and optimizing further did not improve. For MaxCut on Erd\H{o}s-R\'{e}nyi graphs, leveraging VGAE in RQAOA achieved the best ratios over $82.5\%$ of the instances.

\section{Discussion}
\label{conclusion}

In this work, we study different strategies for fixing the parameters of QAOA based on unsupervised learning. We focused on clustering given previous works highlighting the concentration property and exploratory data analysis of the best angles found for MaxCut on Erd\H{o}s-R\'{e}nyi graphs and dense QUBOs. We however use a methodology closer to machine learning by cross-validating compared to related work.
\par 
Furthermore, we demonstrated that these techniques can be leveraged to restrict the number of QAOA circuit calls to small numbers (less than $10$) with a less than $1-2\%$ reduction in approximation ratio on average from the best angles found when cross-validating. We also showed how to compare different clustering strategies and that leveraging instance encodings (by computing features or computing them with a model, in our case a VGAE) for angle setting strategies yields better results than using angle values only. Although the VGAE embedding-based is quite competitive, we recommend using the simpler instance features in practice since the VGAE brings extra computation overhead. For generalization, in regard to the problem scale, both instance features- and VGAE-based clustering approaches manage to retain the performance for unseen problem instances larger than the training set. For dense QUBOs, increasing the clusters is less impactful compared to MaxCut, in which we conjecture that the clusters in QUBO are of large spread and less separable, hindering the performance of the clustering approach in higher dimensions. For both problems, it is necessary to increase the cluster number to retain a good performance when the circuit becomes deeper. 
\par 
From an application perspective, we envision these techniques to be employed in algorithms where QAOA is run on a small part of the problem to solve such as divide-and-conquer \cite{daqqaoa,daqqaoa2} and iterative algorithms \cite{Moussa2021,qls,rqaoa}. Restricting to a few numbers of circuit calls will help decrease the runtime of quantum-featured or quantum-enhanced algorithms, making them closer to competing with classical heuristics. We showcased our approach in the context of Recursive QAOA as hyperparameters under a limited budget (QAOA depth and number of QAOA parameters per iteration limited to $3$), where we were able to achieve a $0.94$ median approximation ratio. With our approaches, we obtain quite comparable performance to the case where we extensively optimize the angles, hence saving numerous circuit calls.

\par 
For future work, other clustering techniques can be studied and extended to predict the angle values by instance in a semi-supervised approach, and for different problem instances. Plus, ML can be used to decide which heuristic to use depending on a given test instance. We also did not apply GNN to the dense QUBOs as graph autoencoders are mostly applied to unweighted graphs. Using VGAE that can reconstruct graph adjacency and node features is then another research direction. Since we use unsupervised methods, we expect the same methodology to be used on noisy hardware. Studying different approaches to resilience under different noisy settings would be also considered of main interest.
Finally, these approaches can be studied within different QAOA-featured algorithms and under different settings (depth of QAOA, number of clusters, Ising instances properties to name a few).

\section*{Acknowledgements}

CM and VD acknowledge support from TotalEnergies. This work was supported by the Dutch Research Council (NWO/OCW), as part of the Quantum Software Consortium programme (project number 024.003.037). This research is also supported by the project NEASQC funded from the European Union’s Horizon 2020 research and innovation programme (grant agreement No 951821).

\bibliography{references}

\begin{thebibliography}{30}
\providecommand{\natexlab}[1]{#1}
\providecommand{\url}[1]{\texttt{#1}}
\expandafter\ifx\csname urlstyle\endcsname\relax
  \providecommand{\doi}[1]{doi: #1}\else
  \providecommand{\doi}{doi: \begingroup \urlstyle{rm}\Url}\fi

\bibitem[Akshay et~al.(2021)Akshay, Rabinovich, Campos, and
  Biamonte]{paramconcentrationfolklore}
Akshay, V., Rabinovich, D., Campos, E., and Biamonte, J.
\newblock Parameter concentrations in quantum approximate optimization.
\newblock \emph{Phys. Rev. A}, 104:\penalty0 L010401, Jul 2021.
\newblock \doi{10.1103/PhysRevA.104.L010401}.
\newblock URL \url{https://link.aps.org/doi/10.1103/PhysRevA.104.L010401}.

\bibitem[Benedetti et~al.(2019)Benedetti, Lloyd, Sack, and Fiorentini]{PQC}
Benedetti, M., Lloyd, E., Sack, S., and Fiorentini, M.
\newblock Parameterized quantum circuits as machine learning models.
\newblock \emph{Quantum Science and Technology}, 4\penalty0 (4):\penalty0
  043001, nov 2019.
\newblock \doi{10.1088/2058-9565/ab4eb5}.
\newblock URL \url{https://doi.org/10.1088/2058-9565/ab4eb5}.

\bibitem[Brand{\~a}o et~al.(2018)Brand{\~a}o, Broughton, Farhi, Gutmann, and
  Neven]{qaoaConcentrates}
Brand{\~a}o, F. G. S.~L., Broughton, M., Farhi, E., Gutmann, S., and Neven, H.
\newblock For {F}ixed {C}ontrol {P}arameters the {Q}uantum {A}pproximate
  {O}ptimization {A}lgorithm's {O}bjective {F}unction {V}alue {C}oncentrates
  for {T}ypical {I}nstances, 2018.
\newblock \url{arXiv:1812.04170}.

\bibitem[Bravyi et~al.(2019)Bravyi, Kliesch, Koenig, and Tang]{rqaoa}
Bravyi, S., Kliesch, A., Koenig, R., and Tang, E.
\newblock Obstacles to state preparation and variational optimization from
  symmetry protection, 2019.

\bibitem[BROYDEN(1970)]{bfgs}
BROYDEN, C.~G.
\newblock The convergence of a class of double-rank minimization algorithms 1.
  general considerations.
\newblock \emph{{IMA} Journal of Applied Mathematics}, 6\penalty0 (1):\penalty0
  76--90, 1970.
\newblock \doi{10.1093/imamat/6.1.76}.
\newblock URL \url{https://doi.org/10.1093/imamat/6.1.76}.

\bibitem[Crooks(2018)]{crooks}
Crooks, G.~E.
\newblock Performance of the quantum approximate optimization algorithm on the
  maximum cut problem, 2018.

\bibitem[Dunning et~al.(2018)Dunning, Gupta, and Silberholz]{dunning}
Dunning, I., Gupta, S., and Silberholz, J.
\newblock What works best when? a systematic evaluation of heuristics for
  max-cut and qubo.
\newblock \emph{INFORMS J. on Computing}, 30\penalty0 (3):\penalty0 608–624,
  aug 2018.
\newblock ISSN 1526-5528.
\newblock \doi{10.1287/ijoc.2017.0798}.
\newblock URL \url{https://doi.org/10.1287/ijoc.2017.0798}.

\bibitem[Farhi \& Harrow(2016)Farhi and Harrow]{qaoaSupremacy}
Farhi, E. and Harrow, A.~W.
\newblock Quantum supremacy through the quantum approximate optimization
  algorithm, 2016.

\bibitem[Farhi et~al.(2014)Farhi, Goldstone, and Gutmann]{QAOA}
Farhi, E., Goldstone, J., and Gutmann, S.
\newblock A quantum approximate optimization algorithm, 2014.

\bibitem[Galda et~al.(2021)Galda, Liu, Lykov, Alexeev, and Safro]{qaoatransfer}
Galda, A., Liu, X., Lykov, D., Alexeev, Y., and Safro, I.
\newblock Transferability of optimal qaoa parameters between random graphs,
  2021.

\bibitem[Guerreschi(2021)]{daqqaoa2}
Guerreschi, G.~G.
\newblock Solving quadratic unconstrained binary optimization with
  divide-and-conquer and quantum algorithms, 2021.

\bibitem[Hansen et~al.(2010)Hansen, Auger, Ros, Finck, and
  Pos{\'{\i}}k]{HansenARFP10}
Hansen, N., Auger, A., Ros, R., Finck, S., and Pos{\'{\i}}k, P.
\newblock Comparing results of 31 algorithms from the black-box optimization
  benchmarking {BBOB-2009}.
\newblock In Pelikan, M. and Branke, J. (eds.), \emph{Genetic and Evolutionary
  Computation Conference, {GECCO} 2010, Proceedings, Portland, Oregon, USA,
  July 7-11, 2010, Companion Material}, pp.\  1689--1696. {ACM}, 2010.
\newblock \doi{10.1145/1830761.1830790}.
\newblock URL \url{https://doi.org/10.1145/1830761.1830790}.

\bibitem[Hinterberger(2009)]{edaref}
Hinterberger, H.
\newblock Exploratory data analysis.
\newblock In \emph{Encyclopedia of Database Systems}, pp.\  1080--1080.
  Springer US, Boston, MA, 2009.
\newblock ISBN 978-0-387-39940-9.
\newblock \doi{10.1007/978-0-387-39940-9_1384}.
\newblock URL \url{https://doi.org/10.1007/978-0-387-39940-9_1384}.

\bibitem[Khairy et~al.(2020)Khairy, Shaydulin, Cincio, Alexeev, and
  Balaprakash]{Khairy2020}
Khairy, S., Shaydulin, R., Cincio, L., Alexeev, Y., and Balaprakash, P.
\newblock Learning to optimize variational quantum circuits to solve
  combinatorial problems.
\newblock \emph{Proceedings of the {AAAI} Conference on Artificial
  Intelligence}, 34\penalty0 (03):\penalty0 2367--2375, April 2020.
\newblock \doi{10.1609/aaai.v34i03.5616}.
\newblock URL \url{https://doi.org/10.1609/aaai.v34i03.5616}.

\bibitem[Kipf \& Welling(2016)Kipf and Welling]{kipf2016variational}
Kipf, T.~N. and Welling, M.
\newblock Variational graph auto-encoders.
\newblock \emph{NIPS Workshop on Bayesian Deep Learning}, 2016.

\bibitem[Kochenberger \& Glover(2006)Kochenberger and Glover]{Kochenberger2006}
Kochenberger, G.~A. and Glover, F.
\newblock A unified framework for modeling and solving combinatorial
  optimization problems: A tutorial.
\newblock In \emph{Multiscale Optimization Methods and Applications}, pp.\
  101--124. Springer US, Boston, MA, 2006.
\newblock ISBN 978-0-387-29550-3.
\newblock \doi{10.1007/0-387-29550-X_4}.
\newblock URL \url{https://doi.org/10.1007/0-387-29550-X_4}.

\bibitem[Lee et~al.(2021)Lee, Saito, Cai, and Asai]{9605323}
Lee, X., Saito, Y., Cai, D., and Asai, N.
\newblock Parameters fixing strategy for quantum approximate optimization
  algorithm.
\newblock In \emph{2021 IEEE International Conference on Quantum Computing and
  Engineering (QCE)}, pp.\  10--16, 2021.
\newblock \doi{10.1109/QCE52317.2021.00016}.

\bibitem[Li et~al.(2021)Li, Alam, and Ghosh]{daqqaoa}
Li, J., Alam, M., and Ghosh, S.
\newblock Large-scale quantum approximate optimization via divide-and-conquer,
  2021.

\bibitem[Lloyd(1982)]{km}
Lloyd, S.
\newblock Least squares quantization in pcm.
\newblock \emph{IEEE Transactions on Information Theory}, 28\penalty0
  (2):\penalty0 129--137, 1982.
\newblock \doi{10.1109/TIT.1982.1056489}.

\bibitem[Moll et~al.(2018)Moll, Barkoutsos, Bishop, Chow, Cross, Egger, Filipp,
  Fuhrer, Gambetta, Ganzhorn, Kandala, Mezzacapo, Müller, Riess, Salis,
  Smolin, Tavernelli, and Temme]{VQE}
Moll, N., Barkoutsos, P., Bishop, L.~S., Chow, J.~M., Cross, A., Egger, D.~J.,
  Filipp, S., Fuhrer, A., Gambetta, J.~M., Ganzhorn, M., Kandala, A.,
  Mezzacapo, A., Müller, P., Riess, W., Salis, G., Smolin, J., Tavernelli, I.,
  and Temme, K.
\newblock Quantum optimization using variational algorithms on near-term
  quantum devices.
\newblock \emph{Quantum Science and Technology}, 3\penalty0 (3):\penalty0
  030503, jun 2018.
\newblock \doi{10.1088/2058-9565/aab822}.
\newblock URL \url{https://doi.org/10.1088/2058-9565/aab822}.

\bibitem[Moussa et~al.(2020)Moussa, Calandra, and Dunjko]{qalgoselection}
Moussa, C., Calandra, H., and Dunjko, V.
\newblock To quantum or not to quantum: towards algorithm selection in
  near-term quantum optimization.
\newblock \emph{Quantum Science and Technology}, 5\penalty0 (4):\penalty0
  044009, oct 2020.
\newblock \doi{10.1088/2058-9565/abb8e5}.
\newblock URL \url{https://doi.org/10.1088/2058-9565/abb8e5}.

\bibitem[Moussa et~al.(2021)Moussa, Wang, Calandra, B{\"a}ck, and
  Dunjko]{Moussa2021}
Moussa, C., Wang, H., Calandra, H., B{\"a}ck, T., and Dunjko, V.
\newblock Tabu-driven quantum neighborhood samplers.
\newblock In Zarges, C. and Verel, S. (eds.), \emph{Evolutionary Computation in
  Combinatorial Optimization}, pp.\  100--119, Cham, 2021. Springer
  International Publishing.

\bibitem[Preskill(2018)]{Preskill2018quantumcomputingin}
Preskill, J.
\newblock Quantum {C}omputing in the {NISQ} era and beyond.
\newblock \emph{{Quantum}}, 2:\penalty0 79, August 2018.
\newblock ISSN 2521-327X.
\newblock \doi{10.22331/q-2018-08-06-79}.
\newblock URL \url{https://doi.org/10.22331/q-2018-08-06-79}.

\bibitem[Sauvage et~al.(2021)Sauvage, Sim, Kunitsa, Simon, Mauri, and
  Perdomo-Ortiz]{flipe}
Sauvage, F., Sim, S., Kunitsa, A.~A., Simon, W.~A., Mauri, M., and
  Perdomo-Ortiz, A.
\newblock Flip: A flexible initializer for arbitrarily-sized parametrized
  quantum circuits, 2021.

\bibitem[{Shaydulin} et~al.(2019){Shaydulin}, {Ushijima-Mwesigwa}, {Safro},
  {Mniszewski}, and {Alexeev}]{qls}
{Shaydulin}, R., {Ushijima-Mwesigwa}, H., {Safro}, I., {Mniszewski}, S., and
  {Alexeev}, Y.
\newblock {Quantum Local Search for Graph Community Detection}.
\newblock In \emph{APS March Meeting Abstracts}, volume 2019 of \emph{APS
  Meeting Abstracts}, pp.\  C42.009, January 2019.

\bibitem[Streif \& Leib(2019)Streif and Leib]{streiftrain}
Streif, M. and Leib, M.
\newblock Training the quantum approximate optimization algorithm without
  access to a quantum processing unit, 2019.

\bibitem[van~der Maaten \& Hinton(2008)van~der Maaten and Hinton]{tsne}
van~der Maaten, L. and Hinton, G.
\newblock Visualizing data using t-sne.
\newblock \emph{Journal of Machine Learning Research}, 9\penalty0
  (86):\penalty0 2579--2605, 2008.
\newblock URL \url{http://jmlr.org/papers/v9/vandermaaten08a.html}.

\bibitem[Wang et~al.(2019)Wang, Zheng, Ye, Gan, Li, Song, Zhou, Ma, Yu, Gai,
  Xiao, He, Karypis, Li, and Zhang]{wang2019dgl}
Wang, M., Zheng, D., Ye, Z., Gan, Q., Li, M., Song, X., Zhou, J., Ma, C., Yu,
  L., Gai, Y., Xiao, T., He, T., Karypis, G., Li, J., and Zhang, Z.
\newblock Deep graph library: A graph-centric, highly-performant package for
  graph neural networks.
\newblock \emph{arXiv preprint arXiv:1909.01315}, 2019.

\bibitem[Zhou et~al.(2018{\natexlab{a}})Zhou, Cui, Hu, Zhang, Yang, Liu, Wang,
  Li, and Sun]{gnnreview}
Zhou, J., Cui, G., Hu, S., Zhang, Z., Yang, C., Liu, Z., Wang, L., Li, C., and
  Sun, M.
\newblock Graph neural networks: A review of methods and applications,
  2018{\natexlab{a}}.

\bibitem[Zhou et~al.(2018{\natexlab{b}})Zhou, Wang, Choi, Pichler, and
  Lukin]{qaoaperf}
Zhou, L., Wang, S.-T., Choi, S., Pichler, H., and Lukin, M.~D.
\newblock Quantum approximate optimization algorithm: Performance, mechanism,
  and implementation on near-term devices, 2018{\natexlab{b}}.
\newblock \url{arXiv:1812.01041}.

\end{thebibliography}
\bibliographystyle{icml2021}

\end{document}